\documentclass[aps,prx,twocolumn,superscriptaddress,noeprint,floatfix,nofootinbib]{revtex4-2}

\usepackage{graphicx}
\usepackage{bm}
\usepackage{amsmath, amssymb}
\usepackage[normalem]{ulem}
\usepackage{physics}
\usepackage{xcolor}
\usepackage{mathtools}
\usepackage[%
  pdfstartview=FitH,%
  breaklinks=true,%
  bookmarks=true,%
  colorlinks=true,%
  anchorcolor=black,%
  citecolor=blue,
  filecolor=black,%
  menucolor=black,%
  urlcolor=blue,%
  linkcolor=blue,%
 ]{hyperref}
\newcommand{\stkout}[1]{\ifmmode\text{\sout{\ensuremath{#1}}}\else\sout{#1}\fi}

\newcommand{\SperpVec}[0]{\vec{{S}}_0^\perp}
\newcommand{\BperpVec}[0]{\vec{{B}}^\perp}
\newcommand{\Sperp}[0]{S_0^\perp}
\newcommand{\Bperp}[0]{B^\perp}

\newcommand{\etamin}[0]{\eta_{\mathrm{min}}}

\def\equationautorefname~#1\null{Eq. (#1)\null}

\newcommand{\figref}[2]{\hyperref[#1]{\autoref*{#1}#2}}
\newcommand{\aref}[1]{\hyperref[#1]{App.~\ref*{#1}}}

\renewcommand\vec{\mathbf}
\newcommand{\Tlya}[0]{T_{\mathrm{lya}}}
\newcommand{\Tth}[0]{T_{\mathrm{th}}}
\newcommand{\Tsep}[0]{T_{\mathrm{sep}}}

\newcommand{\Tphi}[0]{T_\varphi}
\newcommand{\Tmelt}[0]{T_{\mathrm{melt}}}

\newcommand{\Sztot}[0]{S^z_{\mathrm{tot}}}
\newcommand{\HCS}[0]{H_{\mathrm{CS}}}
\newcommand{\HIC}[0]{H_{\mathrm{IC}}}

\newcommand{\pob}[2]{\poissonbracket*{#1}{#2}}

\begin{document}

\title{Confined and deconfined chaos in classical spin systems}

\author{Hyeongjin Kim}
\email{hkim12@bu.edu}
\altaffiliation{equal contribution}
\affiliation{Department of Physics, Boston University, Boston, Massachusetts 02215, USA}

\author{Robin Sch{\"a}fer}
\email{rschaefe@bu.edu}
\altaffiliation{equal contribution}
\affiliation{Department of Physics, Boston University, Boston, Massachusetts 02215, USA}

\author{David M. Long}
\affiliation{Department of Physics, Stanford University, Stanford, California 94305, USA}

\author{Anatoli Polkovnikov}
\affiliation{Department of Physics, Boston University, Boston, Massachusetts 02215, USA}

\author{Anushya Chandran}  
\affiliation{Department of Physics, Boston University, Boston, Massachusetts 02215, USA}
\affiliation{Max-Planck-Institut f\"{u}r Physik komplexer Systeme, 01187 Dresden, Germany}

\date{\today}

\begin{abstract}
Weakly perturbed integrable many-body systems are typically chaotic, and thermal at late times. However, there are distinct relationships between the timescales for thermalization and chaos. The typical relationship is \emph{confined chaos}: when trajectories are still confined to regions in phase space with constant conserved quantities (actions), the conjugate angle variables are already unstable. Chaotic instabilities thus far precede thermalization. 
In a different relationship, which we term \emph{deconfined chaos}, chaotic instabilities and thermalization occur on the same timescale.
We investigate these two qualitatively distinct scenarios through numerical and analytical studies of two perturbed integrable classical spin models: the Ishimori spin chain (confined chaos), and the central spin model with XX interactions (deconfined chaos). We analytically establish (super)-integrability in the latter model in a microcanonical shell. 
Deconfined chaos emerges through the separation of phase space into large quasi-integrable regions and a thin chaotic manifold. The latter leads to chaos and thermalization on the fastest possible timescale, which is proportional to the inverse perturbation strength. This behavior is reminiscent of the quantum SYK models and strange metals.
\end{abstract}
 
\maketitle

\begin{figure*}[t]
    \centering
    \includegraphics[width=0.8\textwidth]{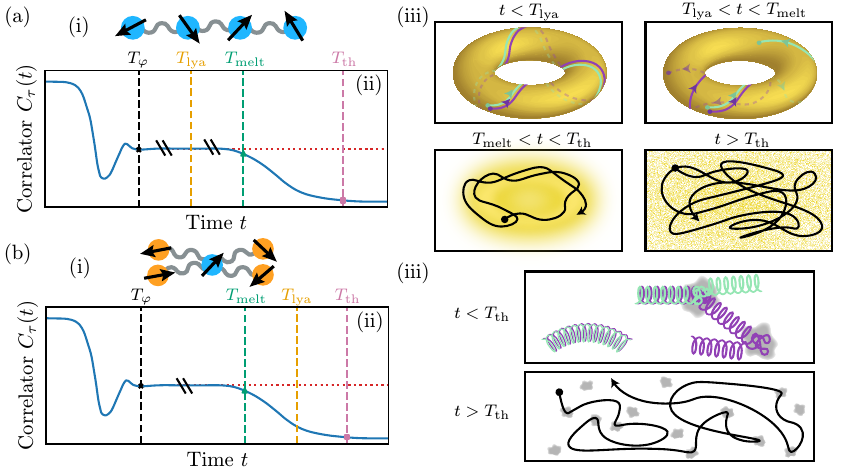}

    \caption[]{Two scenarios for the dynamics of weakly perturbed integrable systems. \emph{Panel (a) corresponds to confined chaos}, which is exhibited by the Ishimori spin chain (i). (ii) is a cartoon of the autocorrelator $C_\tau(t)$ of an observable as a function of time showing four timescales: $\Tphi$ characterizes the timescale on which a trajectory explores a single quasi-invariant torus, $\Tlya$ the timescale of exponential separation of nearby trajectories in phase space, $\Tmelt$ the time at which $C_\tau(t)$ deviates from its late-time value on the torus, and $\Tth$ the time at which $C_\tau(t)$ reaches its thermal value. For confined chaos, $\Tphi \ll \Tlya \ll \Tmelt , \Tth$ so that trajectories rapidly separate within invariant tori before diffusing between tori, as shown in the schematic phase-space cartoons in (iii). \emph{Panel (b) corresponds to deconfined chaos}, which is exhibited by the central spin XX model (i). (ii) shows a cartoon of $C_\tau(t)$ with the hierarchy of timescale $\Tphi \ll \Tlya, \Tmelt, \Tth$. (iii) schematically shows that phase space is extremely inhomogeneous. In the white region, trajectories drift between quasi-invariant tori (which are one-dimensional loops in the model). The gray regions denote the chaotic manifold in which trajectories rapidly separate and correlators relax. Although they appear disconnected for illustrative purposes, they form a connected, thin manifold in phase space.}
    \label{fig:time_scales}
\end{figure*} 

\section{Introduction}
Many physical systems, with length scales ranging from the atomic to the astronomical, are close to integrable. For example, the starting point for describing planetary orbits in our solar system is the (super)-integrable $1/r$-central potential~\cite{LANDAU:1976aa}. At the atomic scale, ultra-cold quantum gases confined to one dimension do not equilibrate even after thousands of collisions, instead exhibiting the dynamics of integrable quantum models for long times~\cite{kinoshita2006quantum,tang2018thermalization,Malvania:2021aa}.  

Mathematically, a Hamiltonian is integrable if it has $L$ (extensive) conserved quantities that are mutually in involution, where $L$ is the number of degrees of freedom~\cite{Ott_2002,strogatz2018nonlinear}.
The conserved quantities can be recast into action variables; their conjugate angle variables increase linearly in time, and the motion of a bounded trajectory in phase space is restricted to a compact $L$-dimensional \emph{invariant torus}. The physical consequences are twofold. First, the dynamics of integrable systems are regular (and thus non-chaotic). Second, the dynamics is non-ergodic; even upon time-averaging, a trajectory stays on its invariant torus, so that the stationary state at late times is described by a generalized Gibbs ensemble (GGE) that in principle accounts for all $L$ conserved quantities.

Upon weakly perturbing an integrable system, it is often observed that chaos asymptotically precedes thermalization (see \figref{fig:time_scales}{a}). Nearby trajectories separate rapidly within tori, so that the precise starting angles are quickly forgotten (\figref{fig:time_scales}{a(iii)} top panels). However, the actions barely change on these timescales, so that the state of the system is well-approximated by a slowly evolving GGE~\cite{kollar_generalized_2011,vidmar_generalized_2016,rigol2007relaxation}. This GGE eventually relaxes to the appropriate Gibbs ensemble on a longer timescale. As the effects of chaos are primarily visible within confined manifolds in phase space, this scenario is called \emph{confined chaos}.

Confined chaos is the basis of Boltzmann kinetics. It is believed to occur in the solar system, wherein numerical simulations predict that the future precise location of a planet in its orbit will be forgotten long before the planet can escape the orbit~\cite{milani_an_1992,milani_stable_1997,laskar_chaotic_2008,winter_short_2010,pereira_confined_2024}. It has also been observed and studied in depth in several model systems~\cite{goldfriend2020quasi}, including in the Fermi-Pasta-Ulam-Tsingou chain~\cite{benettin_the_2013,benettin_fermi_2018,goldfriend2019equilibration}.

Based on empirical studies of a model system, we describe a different scenario in which chaos and thermalization occur asymptotically on the same timescale. We dub this scenario \emph{deconfined chaos}, as trajectories separate in phase space with no reference to angle-like or action-like directions (see \figref{fig:time_scales}{b}). 

While the general conditions for deconfined chaos are not known to us, the essential ingredients in our example are superintegrability and the inhomogeneity of phase space. Superintegrability~\cite{Mishchenko1978superintegrable,Bolsinov2003noncommutative,miller_classical_2013} restricts the motion of the unperturbed trajectory to invariant tori
with dimension less than $L$. A maximally superintegrable system exhibits $2L-1$ independent conservation laws, restricting trajectories to closed loops.
Upon perturbation, trajectories tend to remain close to the tori on short time scales, but slowly drift between distinct tori on longer time scales, leading to a helical motion as illustrated in \figref{fig:time_scales}{b(iii)}.
For weak perturbations, the motion within a torus is typically much faster than the drift, and the helical dynamics can be viewed as a slow drift of the unperturbed torus through phase space.
However, in a small fraction of phase space where the natural timescales of the integrable dynamics are long (gray in \figref{fig:time_scales}{b(iii)}), these tori are completely destroyed by the perturbation. When a drifting trajectory visits this thin chaotic manifold, its dynamics is very sensitive in both the angle-like and action-like directions of phase space, so that all few-body observables thermalize after a few such excursions into the chaotic manifold.

\begin{figure}[!tbp]
    \centering
    \includegraphics[width=\columnwidth]{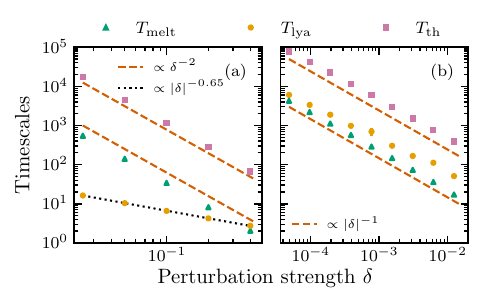}
    \caption[]{Timescales controlling relaxation and Lyapunov instabilities as a function of perturbation strength $\delta$ for (a) the Ishimori spin chain, and (b) the central spin model. Green triangles refer to the melting time $\Tmelt$, orange circles to the Lyapunov time $\Tlya$, and purple squares to the thermalization time $\Tth$. The dashed and dotted lines show relevant power laws. In both models, $\Tmelt$ and $\Tth$ have identical scaling: $\delta^{-2}$ for the Ishimori chain and $\vert\delta\vert^{-1}$ for the central spin. However,  $\Tlya \propto \vert\delta\vert^{-\beta}$ with $\beta\approx 0.65$ for the Ishimori spin chain while $\Tlya \propto \vert\delta\vert^{-1}$ for the central spin model. Models are defined in \autoref{sec:ishimori} and \autoref{sec:centralspin}, respectively. Parameters: $J=1$, $\gamma=3/4$ (only relevant for the central spin model), $L = 45$ in (a), $L=46$ in (b).} \label{fig:real_time_scales}
\end{figure}

We present two case studies, one each for the scenarios of confined and deconfined chaos. In each system, we (i) numerically compute autocorrelators of physical observables in a microcanonical ensemble to access timescales related to thermalization (\(\Tmelt\) and \(\Tth\) in \autoref{fig:time_scales}), and (ii) characterize the sensitivity of a phase-space trajectory to a perturbation through the maximal Lyapunov exponent~\cite{Ott_2002,strogatz2018nonlinear} (this determines the timescale $\Tlya$ in \autoref{fig:time_scales}). We empirically characterize the dependence of these different timescales on the strength of the integrability-breaking perturbation $\vert\delta\vert$ (\figref{fig:real_time_scales}{}), and thus establish confined or deconfined chaos. We also present perturbative arguments to explain the dependence of the various timescales on $\vert\delta\vert$, and build a physical picture of the dynamics.

The case study for confined chaos is the Ishimori chain (\autoref{sec:ishimori}). The Ishimori chain is a classical spin chain with nearest neighbour interactions that supports infinitely long-lived quasi-particles~\cite{ishimori_an_1982,Theodorakopoulos_nontopological_1995,McRoberts_long_2022}.
We find that $\Tlya\propto\vert\delta\vert^{-\beta}$ with $\beta \approx 0.65$, is asymptotically shorter than the timescales associated with thermalization, $\Tmelt,\Tth\propto \delta^{-2}$ (\figref{fig:real_time_scales}{a}). For the studied system sizes, $\Tmelt$ and $\Tth$ also show weak dependence on $L$. However, at much larger sizes, $\Tth$ starts to increase with $L$, as expected from energy diffusion setting the thermalization time, $\Tth\sim L^2/D$ for $D$ being the diffusion constant. The melting time remains approximately system size-independent.

The case study for deconfined chaos  is the central spin XX model (\autoref{sec:centralspin}). This model is known to be integrable when the central spin is quantum with spin $S=1/2$ or $S=1$, where it hosts (exponentially many in $L$) dark states with no dynamics~\cite{villazon_integrability_2020,tang_integrability_2023,villazon_persistent_2020,dimo_strong_2022}. Prior to our work, there have only been cursory investigations of the classical, many-particle limit~\cite{tang_integrability_2023}.

Our first result is that the classical XX model is \emph{maximally superintegrable in a single microcanonical energy shell}. In addition to hosting dark states with no dynamics, the shell is characterized by $2L-1$ independent integrals of motion, which we explicitly compute. Thus, a generic trajectory in this shell traces a closed loop in phase space. Numerically computed Lyapunov exponents are finite in all other energy shells, suggesting that the model is not fully integrable.

Next, we characterize the inhomogeneities in phase space upon perturbation through perturbative arguments and by studying the distributions of various dynamical quantities across phase space. We find that regions in phase space where the dynamics is mean-field-like (either where the mean-field on the central spin is large, or where the field induced by the central spin on every surrounding spin is large) are only weakly perturbed. In contrast, regions of phase space where both fields are weak---so that motion is regular with large periods---are strongly perturbed and give rise to the chaotic manifold. A trajectory encounters this manifold at an average rate of $\vert\delta\vert$. This encounter causes both Lyapunov instabilities and the thermalization of the quantities that are conserved at $\delta=0$. We thus arrive at $\Tlya,\Tmelt, \Tth \propto \vert\delta\vert^{-1}$ (\figref{fig:real_time_scales}{b}). 

Three points are worth noting. First, the emergent timescale in the central spin model is a non-analytic function of the integrability-breaking parameter. Second, the identical scaling of $\Tlya$, $\Tmelt, \Tth$ with $\vert\delta\vert$ prevents the identification of distinct dynamical regimes as $\delta \to 0$. Third are the remarkable similarities between the central spin model and other maximally ergodic models with fast mixing and thermalization, such as the Sachdev-Ye-Kitaev (SYK) model~\cite{Sachdev_2024}.

\section{Ishimori chain} \label{sec:ishimori}
The Ishimori spin chain is an integrable classical model with unbounded nearest neighbour interactions~\cite{ishimori_an_1982}. With periodic boundary conditions on $L$ spins (and identifying $\vb{S}_L$ with $\vb{S}_0$), the Hamiltonian is:
\begin{equation}
    \HIC = J \sum^{L-1}_{j=0} \log(1 + \vb{S}_j \vdot \vb{S}_{j+1}),
\end{equation}
where $J$ is the interaction strength. We set $J = 1$ henceforth\footnote{We note that the equilibrium properties are unstable at low temperatures for this sign choice. However, this does not affect our dynamical results.}.
The model hosts infinitely long-lived quasi-particles~\cite{ishimori_an_1982,Theodorakopoulos_nontopological_1995,McRoberts_long_2022}, and is a special case of the lattice Landau-Lifshitz model~\cite{faddeevHamiltonianMethodsTheory1987,prosen_macroscopic_2013}. It has extensively many conserved quantities, including energy, total $z$-magnetization, and the torsion $\tau$, defined as:
\begin{equation}
    \tau = \sum^{L-1}_{j=0} \frac{\vb{S}_j \vdot (\vb{S}_{j+1} \cross \vb{S}_{j-1})}{(1+\vb{S}_j \vdot \vb{S}_{j+1})(1+\vb{S}_j \vdot \vb{S}_{j-1})}.\label{eq:torsion}
\end{equation}
Dynamics is determined by the usual Hamilton equations \(\dot{A} = \poissonbracket{A}{\HIC}\), where \(\poissonbracket{A}{B}\) is the Poisson bracket of \(A\) and \(B\). The spin variables \(S_j^{x,y,z}\) obey \(\poissonbracket{S_j^a}{S_k^b} = \delta_{jk} \epsilon_{abc} S_j^c\), where \(\epsilon_{abc}\) is the Levi-Civita symbol.

To study the effects of integrability breaking, we consider a perturbation $V$ that adds a spatially inhomogeneous $z$-field to each spin:
\begin{equation}
    V =  \sum_{j=0}^{L-1} h_j S_j^z. \label{eq:ishimori_perturbation}
\end{equation}
The field strengths $h_j \in [-1,1]$ are generated pseudo-randomly using Sobol sampling. 
We use Sobol sampling, as opposed to random sampling, as the exact distribution of the $h_j$ does not affect any of our results. Thus, we fix an unstructured but reproducible realization. The perturbed Hamiltonian $H = \HIC + \delta\, V$ (we take $\delta \geq 0$) conserves energy and total $z$-magnetization, but does not conserve the torsion \(\tau\). Fixing both energy $E$ and total $z$-magnetization $\Sztot=\sum_j S_j^z$ to zero specifies a microcanonical shell, to which we restrict our numerics.

\subsection{Decay of autocorrelators} \label{subsec:ishimori_relaxation}

\begin{figure}[t]
    \centering
    \includegraphics[width=\columnwidth]{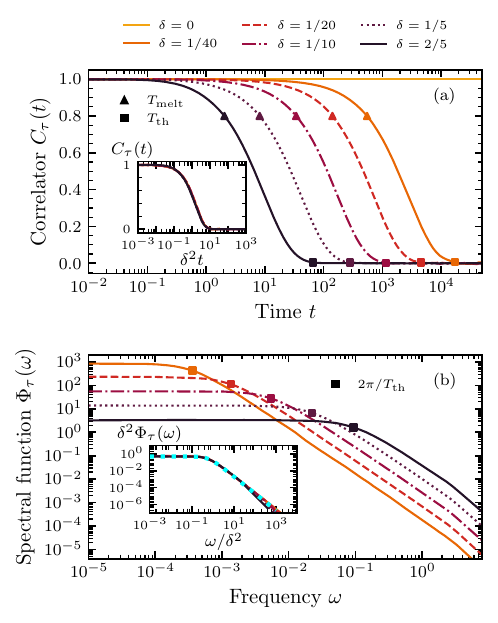}
    \caption[]{Dynamics of the torsion $\tau$ as defined in \autoref{eq:torsion} in the Ishimori spin chain with system size $L = 45$ at varying perturbation strengths $\delta$. Panels (a) and (b) show the autocorrelators and spectral functions of $\tau$, respectively. The triangles and squares mark the melting time $\Tmelt$ and thermalization time $\Tth$, respectively. In (a), the correlator is constant, $C_{\tau}(t) = 1$, at $\delta = 0$ (shown in light orange). The insets show rescaled data, indicating $\Tmelt, \Tth \propto \delta^{-2}$. The inset in (b) further shows that the Lorentzian function (shown as blue dotted line) is a good fit to $\Phi_{\tau}(\omega)$.} \label{fig:ishimori_cspec}
\end{figure} 

To investigate the process of thermalization, we study the autocorrelators of observables and their spectral functions as a function of perturbation strength $\delta$, as shown in \autoref{fig:ishimori_cspec}. Specifically, we consider \textit{normalized} autocorrelators,
\begin{equation}
    C_O(t) = \frac{\expval{O(t)O(0)}}{\expval{O(0) O(0)}},\label{eq:correlation}
\end{equation}
where $\expval{\ldots}$ refers to the phase-space average (mean) over the microcanonical shell. We also examine the spectral function, which is defined as the Fourier transform of the autocorrelator:
\begin{equation}
    \Phi_O(\omega) = \int^\infty_{-\infty} \frac{\dd{t}}{2\pi} e^{i\omega t} C_O(t) . \label{eq:spectral_function}
\end{equation}
See Ref.~\cite{lim2024defining} or \aref{app:corr_spec} for specific details in computing the correlators and spectral functions.

We consider the torsion $\tau$ from \autoref{eq:torsion} as our observable $O$ with $\tau(t)=\tau(0)$ at $\delta=0$. For a conserved quantity, the relaxation time at the integrable point, denoted $T_\varphi$ in \figref{fig:time_scales}{}, plays no role, and the \emph{prethermal plateau}~\cite{langen2016prethermalization,mori2018thermalization} (shown as a red horizontal dotted line in \figref{fig:time_scales}{}) is situated at $C_\tau(t) = 1$. This choice simplifies the extraction of the remaining timescales and homes in on the effects of integrability breaking. For non-conserved quantities, $\Tphi \propto \delta^0$ sets the timescale for the initial relaxation to the prethermal plateau\footnote{Note that $T_\varphi$ generally increases with the system size for local integrable models. As we are interested in the onset of thermalization at a fixed $L$, all other $\delta$-dependent scales exceed $\Tphi$ for small enough \(\vert\delta\vert\)}. More details can be found in \aref{app:non_conserved_observables}. 

\figref{fig:ishimori_cspec}{a} and \figref{fig:ishimori_cspec}{b} show the autocorrelator and spectral function at different perturbation strengths for the Ishimori spin chain with $L = 45$. The correlator decays from its initial value at $C_\tau(t) = 1$ and eventually relaxes to its microcanonical value $\expval{\tau}^2/\expval*{\tau^2}=0$. We extract the melting time $\Tmelt$ and thermalization time $\Tth$.

The melting time $\Tmelt$ is extracted as the time at which $C_\tau(t) = 0.8$ and marked by triangles in \figref{fig:ishimori_cspec}{a}. To illustrate its dependence on $\delta$, we plot the autocorrelator versus re-scaled time in the inset, which indicates the scaling $\Tmelt \propto \delta^{-2}$. 

The thermalization time $\Tth$ is extracted using the spectral function, which broadens near $\omega = 0$ as $\delta$ increases. Specifically, the divergent spectral weight at $\omega=0$ at the integrable point is redistributed to other frequencies at any nonzero $\delta$.
Assuming that the autocorrelator exponentially relaxes to the microcanonical value at the longest times, the spectral function in the neighborhood of $\omega=0$ can be fit to a Lorentzian,
\begin{equation}
    \Phi_\tau(\omega) \approx \frac{1}{\pi} \frac{\Gamma}{\omega^2 + \Gamma^2} ,\label{eq:spectral_lorentzian}
\end{equation}
with width $\Gamma$ that defines the thermalization time as $\Tth \equiv 2\pi/\Gamma$. The inverse Fourier transform of the Lorentzian in \autoref{eq:spectral_lorentzian} gives $e^{-\Gamma \abs{t}}$ for the form of $C_\tau(t)$ at late times. Then, $\Tth$ (marked by squares in \figref{fig:ishimori_cspec}{a}) sets the time at which the correlator has decayed to approximately $e^{-2\pi} \approx 0.2\%$ of its original value. 

Similar to the melting time, the thermalization time $\Tth$ is proportional to ${\delta^{-2}}$.
The inset in \figref{fig:ishimori_cspec}{b} shows a complete scaling collapse of the spectral function upon rescaling the $x$-axis as $\omega\to\omega/\delta^2$ and the $y$-axis as $\Phi_\tau\to \delta^2 \Phi_\tau$. This scaling collapse implies that $\Phi_\tau(\omega)\approx \Gamma^{-1}\phi_\tau(\omega/\Gamma)$ with $\Gamma\propto \delta^2$ and some dimensionless scaling function $\phi_\tau(x)$, which is well approximated by a pure Lorentzian function (shown as a blue dotted line). 

Thus, the entire thermalization process is determined by a single timescale ($\propto \delta^{-2}$) set by $\Tmelt$ or $\Tth$. This is consistent with standard time-dependent perturbation theory used in deriving kinetic equations for approximate integrals of motion and Fermi's Golden Rule (FGR), which is equivalent to the relaxation time approximation for these equations. Within these approaches, the transition rates between quasi-stationary GGE configurations are proportional to
\begin{align}\label{eq:FGR}
    \Gamma_{\mathrm{pert}} \propto \frac{1}{L}\delta ^2 \expval*{V^2}_{cc} \Phi_{V}(0),
\end{align}
where $\Phi_V$ is the spectral function of the perturbation (note that $\Phi_V$ is normalized as defined in \autoref{eq:spectral_function}) and $cc$ refers to the connected correlator. {The factor of $1/L$ comes from the kinetic constraint that an extensive integral of motion can only change by an intensive amount in each transition event. This factor is compensated for by the connected correlator $\expval*{V^2}_{cc}\propto L$.} Thus, the corresponding melting time $T_{\rm melt}$ and the thermalization time $T_{\rm th}$ of the initially conserved operators are both proportional to $( \delta^2 \Phi_{V}(0))^{-1}$. 

Notice that these times can significantly differ from each other: $T_{\rm melt}$ depends on the spectral function $\Phi_V(0)$ computed with the initial GGE ensemble, while $T_{\rm th}$ is defined by the equilibrium, microcanonical spectral function. Therefore, we do not expect that the relaxation of $C_O(t)$ will be generically described by a single timescale. If, however, the initial GGE distribution is sufficiently close to the microcanonical ensemble, as is the case for our studies, the relaxation rate remains approximately time-independent. Crucially, the perturbation theory prediction is well controlled when the spectral function of the perturbation is finite at low frequency, \(\Phi_{V}(0) < \infty\). This is the case for the perturbation \( V\) in the Ishimori chain as shown in \aref{app:non_conserved_observables}. 

\subsection{Divergence of trajectories}\label{ssec:Ishimori_rates}

The melting and thermalization times \(\Tmelt\) and \(\Tth\) only provide us with information about actions---the approximate integrals of motion. Thus, they are completely insensitive to angle variables. In principle, one can study those in a similar fashion, but it is generally more difficult as explicit analytic expressions for the angle variables are not readily available. To circumvent this problem, we analyze the Lyapunov time $\Tlya$, which characterizes the earliest instability occurring in either action or angle variables. Formally, the Lyapunov time is defined as the inverse of the maximal Lyapunov exponent $\lambda$.

The largest Lyapunov exponent $\lambda$ characterizes the exponential separation of nearby trajectories averaged over the entire phase space. It is often used as a measure of chaos in non-linear dynamical systems~\cite{Ott_2002,skokos_the_2010,strogatz2018nonlinear,benettin1980,benettin1976kolmogorov,strogatz2018nonlinear}. The exponent $\lambda$ is computed by evolving two nearby trajectories with initial conditions $\vb{S}(0)$ and $\widetilde{\vb{S}}(0)$ and measuring the growth of the distance $d(t) = \norm*{\vb{S}(t) - \widetilde{\vb{S}}(t)}_2$ as a function of time $t$:
\begin{equation}
    \lambda = \lim_{T\to \infty}\lim_{d_0 \to 0} \frac{1}{T} \log{\left(\frac{d(T)}{d_0}\right)}.
    \label{eq:lyapunov}
\end{equation}
The timescale $\Tlya$ is defined by:
\begin{equation}
    \Tlya = \lim_{T\to \infty}\lim_{d_0 \to 0} \frac{T}{\log{\left(\frac{d(T)}{d_0}\right)}}.
    \label{eq:Tlya}
\end{equation}

If the system is ergodic or mixing, then the Lyapunov time as defined above does not depend on the initial conditions. However, it can take a very long time $T$ for $\Tlya$ to converge to the correct asymptotic value. To circumvent this issue, we choose a large but finite $T$ and average the Lyapunov time over the microcanonical shell of initial conditions. Specifically, we choose a small initial distance $d_0$ and define a large cutoff growth factor, \(a\), from which we compute the \emph{separation times} $\Tsep$ for each initial point $\vec{S}_j(0)$, 
\begin{equation}
    T_{\mathrm{sep}}(\vec{S}_j) = \frac{T_{\mathrm{final}}(\vec{S}_j)}{\log{\left(\frac{d(T_{\mathrm{final}}(\vec{S}_j))}{d_0}\right)}}.
    \label{eq:tsep}
\end{equation}
Here, $T_{\mathrm{final}}(\vec{S}_j)$ is defined as the time when the separation $d(t)$ has grown from $d(0)=d_0$ to $d(T_{\mathrm{final}}(\vec{S}_j))=ad_0$ for the $j$th sample $\vec{S}_j$ within the microcanonical shell. The sample mean of the separation time for a sample of \(n \gg 1\) initial configurations approximates $\Tlya$:
\begin{align}
    \Tlya\approx\expval{\Tsep}=\frac{1}{n}\sum_{j=1}^n {\Tsep(\vec{S}_j)}=\frac{1}{\log(a^n)}\sum_{j=1}^n {T_{\mathrm{final}}(\vec{S}_j)}.
\end{align}
The sum over the ${T_{\mathrm{final}}(\vec{S}_j)}$ values approximates the limit $\lim_{T \to \infty}$ in \autoref{eq:Tlya}, as it effectively represents the total time $\sum_j {T_{\mathrm{final}}(\vec{S}_j)}$ required for the distance to grow to $a^n$. We have verified that our values of $d_0$ (small), $a$ (large), and $n$ (large) are chosen such that the resulting estimate for $\Tlya$ is well converged.

Within our simulations, we have chosen both trajectories to initially start within the same invariant torus (in the integrable limit). We obtain a second trajectory within the same quasi-invariant torus by evolving under integrable dynamics for a short time proportional to $d_0$. This avoids a linear separation of the trajectories proportional to the initial distance due to a mismatch in the action variables.
A comparison to an alternative approach using out-of-time-order correlators (OTOC)~\cite{Larkin:1968aa,rozenbaum_lyapunov_2017} is given in \aref{app:lyapunov_otocs}.

\begin{figure}[t]
    \centering
    \includegraphics[width=\columnwidth]{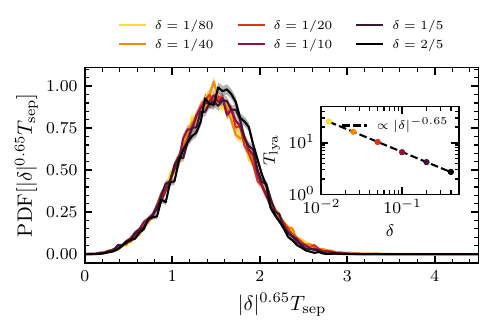}%
    \caption[]{Distribution of the rescaled separation time $\vert\delta\vert^{\beta} \Tsep$ in the Ishimori spin chain (with system size $L = 45$) at varying perturbation strengths $\delta$ each for $10,000$ initial spin configurations. The inset shows the mean of the distribution, $\Tlya$, as a function of $\delta$. The parameter $\beta \approx 0.65$ is obtained by fitting $\Tlya \propto \abs{\delta}^{-\beta}$ using bootstrapping, whose fit is shown in the inset.
    \label{fig:ishimori_tlya}}
\end{figure}

The distribution of \(\Tsep\) over phase space for different perturbation strengths $\delta$ is shown in \autoref{fig:ishimori_tlya}.
The scaling of the entire distribution follows a simple power law with respect to $\vert\delta\vert$, leading to a scaling collapse when rescaling the distribution as $\vert\delta\vert^\beta \Tsep$ with $ \beta \approx 0.65$. Consequently, the mean, corresponding to $\Tlya$, scales as $\vert\delta\vert^{-\beta}$, as shown in the inset. 
The collapse in \autoref{fig:ishimori_tlya} suggests that the perturbation breaks integrability in a similar manner across phase space. 

Similar scalings ($\beta \approx 2/3$) have been observed for classical and quantum kicked rotors that are perturbed by Gaussian noise~\cite{lam2014stochastic,goldfriend2020quasi}.  The approximate equality of \(\beta\) between the two models is, however, possibly accidental, as other models studied in the literature have different values of $\beta$~\cite{falcioni_ergodic_1991,mulansky_strong_2011,merab_lyapunov_2022,wijn_largest_2012}.

\subsection{Confined chaos} \label{ssec:confined_chaos}

The timescale at which trajectories separate in phase space is asymptotically shorter than the timescale at which the system thermalizes: 
\begin{equation}
    \Tmelt, \Tth \propto \delta^{-2}\gg \Tlya\propto \vert\delta\vert^{-\beta} \quad\text{with }\beta\approx 0.65
\end{equation}
for $\delta\rightarrow 0$ (also shown in \figref{fig:real_time_scales}{a}). This phenomenon is the norm rather than an exception in nearly integrable systems and is known as confined chaos~\cite{goldfriend2019equilibration,goldfriend2020quasi}. Canonical examples are the weakly interacting atomic gas and planetary orbits in the Solar System~\cite{milani_an_1992,milani_stable_1997,laskar_chaotic_2008,winter_short_2010,pereira_confined_2024}. Trajectories initially separate within the quasi-invariant tori, which begin to melt (and thus autocorrelators of their associated actions begin to decay) at much later times. This scenario usually justifies FGR and various Markovian approximations, including kinetic approaches for gases, where one only writes equations of motion for the actions (such as the momenta of atoms in a colliding gas) while assuming that the corresponding angles are completely random.
In real many-body systems, Lyapunov exponents are hard or even impossible to probe because of quantum or statistical fluctuations, which prevent us from creating two copies of the system with nearly identical initial conditions. Nevertheless, confined chaos can be understood as a two stage relaxation process where angle variables equilibrate at a timescale faster than the action variables.

Based on the asymptotically different timescales, the Ishimori spin chain exhibits four distinct dynamical regimes as illustrated in \figref{fig:time_scales}{a(iii)}:
\begin{itemize}
    \item [(i)] \textit{Regular GGE} ($\Tphi < t < \Tlya$): Trajectories remain perturbatively close to those of the integrable model. 
    \item [(ii)] \textit{Chaotic GGE} ($\Tlya < t < \Tmelt$): Nearby phase-space trajectories have exponentially diverged, but the quasi-invariant tori have not yet melted. A comprehensive picture of the dynamics in phase space in this regime can be found in Ref.~\cite{goldfriend2020quasi}. The temporal extent of the chaotic GGE regime grows as $\delta \to 0$.
    \item [(iii)] \textit{Melting} ($\Tmelt < t < \Tth$):  The quasi-conserved quantities begin to melt as the (already) chaotic trajectories start to explore different tori. In the case of torsion, its autocorrelator $C_\tau(t)$ decays exponentially as a function of time.
    \item [(iv)] \textit{Thermalization} ($t > \Tth$): The exponential decay of $C_\tau(t)$ continues until thermal equilibrium is reached at $\Tth$, after which $C_\tau(t)$ remains constant, up to fluctuations which are suppressed by $L$.
\end{itemize}

\subsection{System-size dependence of thermalization}\label{ssec:ishimori_systemsize}

We conclude our discussion of the Ishimori spin chain by examining the system-size dependence of the thermalization time $\Tth$, with numerical evidence provided in \aref{app:finite_size}.

We expect the Ishimori chain to be diffusive. Therefore, the thermalization time $\Tth$ of a generic local observable is expected to be system-size dependent. To be specific, a local observable with non-vanishing overlap with a conserved quantity in the perturbed model (i.e., energy or total magnetization) generally exhibits power-law decay of the autocorrelator, which emerges as a low-frequency power law in the spectral function (see \aref{app:finite_size} for observable $S_0^z$). For example, in our one-dimensional system, we find diffusive scaling of $\Phi_{S_0^z}(\omega) \sim 1/\sqrt{\omega}$. This power-law scaling ends at the Thouless frequency (inverse thermalization time), which scales as $\omega_{\rm Th}\sim D(\delta)/L^2$, where $D(\delta)$ is the diffusion constant. Thus, the thermalization time of a local observable is generally system-size dependent ($\Tth \propto L^2$)\footnote{Note that the melting time, indicating the onset of the destruction of the tori, is independent of the observable's overlap with conserved quantities and thus remains system-size independent.}.

However, we find that the thermalization time of torsion $\tau$ is system-size independent because it has a vanishing overlap with the conserved quantities of the perturbed model, and thus does not exhibit a power-law decay of the correlator at late times.

Even for observables which overlap with conserved quantities, observing power-law relaxation of the autocorrelator requires a system size \(L\) which is sufficiently large compared to \( \sqrt{D(\delta)} \vert\delta\vert^{-1}\) such that $L^2 / D(\delta) \gg \Tth $. Note that $D(\delta)$ usually diverges as $\delta \to 0$~\cite{ljubotina2017spin,nardis2021stability}, the general mechanism for which is yet to be explained. Taking the limit $\delta \to 0$ at a fixed system size, there is no  asymptotic diffusive relaxation even for observables with non-vanishing overlaps with conserved quantities (see \aref{app:non_conserved_observables} for more details); instead, autocorrelators exponentially decay. This is because quasi-particles bounce back and forth between the system boundaries many times before eventually decaying, so that the diffusion of energy does not set the time-scale for reaching thermal equilibrium. In the opposite limit of fixed $\delta$ and $L\to\infty$, the system first locally thermalizes and then conserved densities evolve in time according to the diffusion equation or, more generally, according to hydrodynamics. In this case, the global thermalization time \emph{does} depend on the system size. For weak integrability breaking, studying this regime would require much larger system sizes than we analyze here.

\section{Central spin model with XX interactions} \label{sec:centralspin}
Central spin models provide an idealized description of interactions between a central degree of freedom and a mesoscopic environment of surrounding spins, as depicted in \figref{fig:time_scales}{b(i)}. They model the interactions between nitrogen-vacancy centers and nuclear spins
in diamond~\cite{schwartz_robust_2018, London:2013}, and the hyperfine interaction between electronic and nuclear spins in quantum
dots~\cite{Urbaszek:2013, hanson_spins_2007}. Large families of such models are known to be integrable~\cite{dukelsky_colloquim_2004,yuzbashyan_solution_2005,dobrzyniecki_quantum_2023}; they thus provide a controlled setting within which to study relaxation dynamics upon an integrability-breaking perturbation. 

We focus on the classical XX central spin model, in which the central spin interacts with its environment resonantly through XX interactions. The Hamiltonian is,
\begin{equation}
    \HCS = J\sum^{L-1}_{j=1}g_j(S_0^x S_j^x + S_0^yS_j^y),\label{eq:CS}
\end{equation}
where $0$ indexes the central spin, $j=1, \ldots, L-1$ indexes each environment spin, and $g_j = 1+ \gamma g_j^\prime$ with $g_j^\prime \in [-1,1]$, which are obtained using Sobol sampling. Again, we use Sobol sampling to obtain an unstructured but reproducible realization. Above, $J$ sets the overall strength of the interaction between the central spin and each environment spin, while $\gamma$ sets the strength of the deviation from $J$. We set $J = 1$ and $\gamma=3/4$ in our simulations.

The unperturbed Hamiltonian can be expressed in terms of a collective mean field \( \vb{B} \) generated by the environmental spins:
\begin{align}
    \vb{B} = \sum_{j>0} g_j \vec S_j.
\end{align}

The central spin Hamiltonian \( \HCS \) then takes the form of an effective two-body interaction between the central spin \( \vb{S}_0 \) and the mean field \( \vb{B} \): 

\begin{equation}
    \HCS = J\,{\bf \Sperp}\cdot {\bf \Bperp}= J\, \cos(\varphi_0 - \varphi_B)\, S_0^{\perp} B^{\perp}, \label{eq:MF}
\end{equation}
where \( \varphi_0 \) and \( \varphi_B \) are the azimuthal angles of the central spin and mean field, respectively, and \( S_0^{\perp} = \sqrt{(S_0^x)^2 + (S_0^y)^2} \), \( B^{\perp} = \sqrt{(B^x)^2 + (B^y)^2} \) denote the projections onto the $xy$-plane relevant to the XX interaction. 

We investigate chaos and thermalization in the central spin XX model when perturbed by a spatially inhomogeneous $z$-field. The field strengths, $h_j \in [-1,1]$, are also generated using Sobol sampling:
\begin{equation}
    V =   \sum_{j=1}^{L-1} h_j S_j^z. \label{eq:centralspin_perturbation}
\end{equation}
The sum over $j$ excludes the central spin $\vb{S}_0$.

The perturbed central spin Hamiltonian $H=\HCS+\delta\, V$ has a conserved total energy $E$ and total magnetization $\Sztot$. All simulations in the study are carried out within the microcanonical shell obtained by setting $E=0$ and $\Sztot=0$.

\subsection{Superintegrability at zero energy}
We prove that the XX central spin model at $E=0$ is maximally superintegrable by identifying $2L-1$ integrals of motion, which confine the motion to a one-dimensional closed curve in phase space. A Hamiltonian system is considered superintegrable when the number of independent integrals of motion exceeds $L$, which is the number of conserved quantities required to guarantee integrability (assuming they are in involution)~\cite {Mishchenko1978superintegrable,Bolsinov2003noncommutative,miller_classical_2013}. If there are more than \(L\) functionally independent integrals of motion, they cannot all be in involution. Nevertheless, each integral reduces the dimension of the phase space surface explored by a trajectory. In maximally superintegrable systems with $2L-1$ integrals, trajectories in the $2L$-dimensional phase space are confined to one-dimensional curves. A prominent example of a maximally superintegrable model (without the energy restriction $E=0$) is the isotropic all-to-all Heisenberg model~\cite{kittel_development_1965,magyari_integrable_1987,liu_infinte_1990}. Another prominent and well-known example is the Kepler problem, which describes a three-dimensional particle in a $1/r$ potential.

The proof of superintegrability in the \(E=0\) central spin model is presented in  \aref{app:regular_CS}. Here, we give a brief summary of that analysis. The primary observation is that $S_0^{x}/S_0^{y} = -B^y/B^x$ is conserved at zero energy, as can be directly checked from the equations of motion. Thus, the angles $\varphi_0$ and $\varphi_B$ are conserved in time modulo $\pi$ and the central spin is confined to a fixed great circle.
In turn, each environmental spin perceives a time-dependent magnetic field pointing along/against $\bf\Sperp$. In this way, the angle between each environment spin and the central spin's projection to the \(xy\)-plane is conserved in time. This establishes $L$ conserved quantities, where each has support on at most two spins. The $L-2$ additional conservation laws are a consequence of each environment spin being subjected to the same time-dependent magnetic field proportional to  $\Sperp(t)$. Together with the total magnetization, we arrive at $2L-1$ functionally independent conserved quantities. The energy is not an additional conserved quantity, as it can be expressed as a function of the \(2L-1\) quantities we have identified.

Spin dynamics is described by:
\begin{align}
    \vec{\dot{S}_0} = J \vec{S_0}\cross {\bf \Bperp} \text{ and } \vec{\dot{S}_j} = Jg_j\, \vec{S_j}\cross{\bf \Sperp}. \label{eq:EOM_CS}
\end{align}
Since the {directions of the vectors $\bf \Bperp$ and $\bf \Sperp$ are conserved in time (modulo $\pi$)}, the central spin precesses around the mean field's $\bf \Bperp$ and the environment spins precess around the central spin's projection $\bf\Sperp$. Note that the energetic constraint requires $\bf \Bperp\cdot\bf \Sperp=0$. The respective strengths $B^\perp$ and $S_0^\perp$ are not conserved, rendering 
\autoref{eq:EOM_CS} as a nonlinear system of equations which we solve numerically. 

We make two observations.
(i) Solutions to \autoref{eq:EOM_CS} are periodic (see \aref{app:regular_CS}). 
This is consistent with the one-dimensional nature of trajectories in maximally superintegrable systems. We can assign a period $T$ to each orbit and an angle $\alpha$ characterizing the position on the orbit. Note that \(\alpha\) is not an angle in the action-angle framework as \(\alpha\) is not linearly evolving in time. 
(ii) The precession frequencies become small when $S_0^\perp$ and $B^\perp$ are simultaneously close to zero. This regime is important in the next section. To characterize this regime, we introduce the parameter 
\begin{align} 
    \eta = \sqrt{(S_0^{\perp})^2 + (B^{\perp})^2}. \label{eq:eta}
\end{align}
The condition $\eta = 0$ defines a $(2L-4)$-dimensional submanifold in phase space. States on this manifold are frozen because the right-hand sides of \autoref{eq:EOM_CS} vanish when $S_0^\perp = 0$ and $B^\perp = 0$. These states are analogues of the dark states in the quantum version of the model~\cite{villazon_persistent_2020,villazon_integrability_2020,wu_separable_2020,tang_integrability_2023,nadai_integrability_2024,tonder_dark_2025}. Note that it is not possible to define the conserved quantities for $\eta=0$.

To perform numerical analysis, we select an integral of motion associated with the central spin:
\begin{align}
    \label{eq:centralspin_observable}
    \tau_0=2\frac{S_0^xS_0^y}{ ( S_0^{\perp} )^2}=\sin{2\varphi_0}.
\end{align}
The energetic constraint \(E=0\) relates $\tau_0$ to the analogous quantity for the mean field:
\begin{align}
    \label{eq:tau_B}
    \tau_B=2\frac{B_0^xB_0^y}{ ( B^{\perp} )^2}=\sin{2\varphi_B}.
\end{align}
We emphasize that $\tau_0$ and $\tau_B$ have discontinuities when $B^{\perp}$ and $S_0^{\perp}$ vanish, respectively. However, they remain bounded, and further are constant along trajectories, even if trajectories pass through \(S_0^{\perp}=0\) or \(B^{\perp} = 0\).

Numerical simulations away from the microcanonical zero-energy shell reveal finite Lyapunov exponents and anomalously slow relaxation dynamics typical for classical models with mixed phase space or strongly disordered models. This refutes the conjectures of Ref.~\cite{tang_integrability_2023} stating that the classical XX model is integrable.

\subsection{Effect of integrability breaking}\label{ssec:inhomogeneity}

\begin{figure}[t]
    \centering
    \includegraphics[width=\columnwidth]{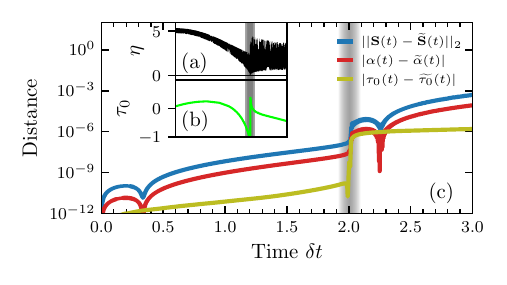}
    \caption{A typical trajectory which traverses the chaotic manifold around time $t \approx 2\vert\delta\vert^{-1}$ (gray shaded region). Panel (a) displays the instability parameter $\eta$ defined in \autoref{eq:eta}; panel (b) shows the integral of motion $\tau_0$ from \autoref{eq:centralspin_observable}; and panel (c) compares two nearby trajectories, $\mathbf{S}$ and $\widetilde{\mathbf{S}}$. The blue curve represents their Euclidean distance in spin variables (initialized at $10^{-12}$), the red curve shows the angular degree of freedom $\alpha$ characterizing the one-dimensional orbit for $\delta=0$, and the yellow curve tracks the distance between the two trajectories in terms of the integral of motion $\tau_0$. We compute $\tau_0$ when it penetrates the $xy$-plane to avoid poles.}
    \label{fig:example}
\end{figure}

The perturbation $V$ breaks superintegrability in a highly inhomogeneous manner across phase space, leading to quantitatively distinct dynamics in different regions. In regions of phase space where $\eta$ is smaller or comparable to the perturbation strength $\delta$, trajectories rapidly separate in all coordinates, and quasi-conserved quantities change rapidly as a function of time. We refer to the thin region of phase space where $\eta \lesssim \vert\delta\vert$ as the \textit{chaotic manifold}. In contrast, when $\eta$ is large, trajectories closely follow their $\delta=0$ closed curves. We refer to this region as \textit{quasi-integrable}. Most of the phase space lies within the quasi-integrable domain, and we empirically find that the chaotic manifold is traversed only intermittently, at an average frequency proportional to $\vert\delta\vert$.

In the remainder of this subsection, we examine the dynamics within each region. The implications for the decay of the autocorrelator and the behavior of Lyapunov exponents and the respective data are discussed in \autoref{ssec:decay_AC} and \autoref{ssec:divergence}. A typical trajectory passing through a chaotic manifold is shown in \autoref{fig:example}. \aref{app:supp_CS} provides further data supporting our observations.

\paragraph*{Quasi-integrable regime.}
Within the quasi-integrable regime, either $S_0^{\perp}$ or $B^{\perp}$ (or both) exceed the perturbation strength $\vert\delta\vert$.
In this case, dynamics is close to that of \autoref{eq:EOM_CS}, and the central and environmental spins jointly precess as in the integrable case. However, there are two consequences of $\delta \neq 0$: (i) quasi-conserved quantities exhibit a steady drift at a rate proportional to $\vert\delta\vert$, and (ii) the separation of nearby trajectories is predominantly determined by the angle coordinate describing motion along the closed orbit at $\delta=0$.
We focus on (i), the drift of the conserved quantities. Point (ii) is a comparatively straightforward empirical observation.

\paragraph*{$\rm (i)$} Quasi-conserved quantities drift linearly with a velocity proportional to $\abs{\delta}$.
To demonstrate this, consider the equations of motion for the conserved quantities \(\tau_0\) and \(\varphi_0\) on the \(E=0\) shell. The perturbation alters the equation of motion of the environment spins from \autoref{eq:EOM_CS} to
\begin{align}
   \vec{\dot{S}_j} = J \vec{S_j}\cross\left[g_j{\bf \Sperp}+\delta\begin{pmatrix} 0 \\ 0 \\h_j\end{pmatrix} \right]. \label{eq:EOM_CS_Pert}
\end{align}

The equations of motion for the central spin observables are formally unchanged
\begin{align}
     \dot{\tau}_0 &= -2\frac{(S_0^x)^2-(S_0^y)^2}{\left( S_0^{\perp}\right)^4}S_0^z \HCS \text{, and } \dot \varphi_0=-\frac{S_0^z}{\left( S_0^{\perp}\right)^2} \HCS.\label{eq:dot_tau_primitive}
\end{align}
However, \(\HCS\) is no longer a conserved quantity of the full dynamics.
(The fact that the right-hand side is proportional to \(\HCS\) is what makes these quantities conserved on the \(E=0\) shell for \(\delta=0\).)
With nonzero \(\delta\), we can use the constraint \(E = \HCS(t) + \delta V(t)\) to re-express these equations on the \(E=0\) microcanonical shell in a way which explicitly shows their dependence on \(\delta\),
\begin{align}
     \dot{\tau}_0 &= 2\delta\frac{(S_0^x)^2-(S_0^y)^2}{\left( S_0^{\perp}\right)^4}S_0^zV \text{, and } \dot \varphi_0=\delta\frac{S_0^zV}{\left( S_0^{\perp}\right)^2}.\label{eq:dot_tau}
\end{align}

From \autoref{eq:dot_tau}, we can deduce the linear drift. Consider, for example, the azimuthal angle $\varphi_0$. 
Let \(T\) be the period of the unperturbed motion for a nearby orbit on the \(\expval{\HCS} = 0\) shell in the unperturbed model. The total change in \(\varphi_0\) over time \(T\) is given by the integral;
\begin{equation}
    \label{eq:Delta_Phi}
    \Delta\varphi_0={\delta}\int_0^T \frac{S_0^z(t) V(t)}{\left( S_0^{\perp}(t)\right)^2} \dd t.
\end{equation}
If the period \(T\) is short compared to \(\vert\delta\vert^{-1}\), as happens when \(\eta \gg \vert\delta\vert\), and additionally \(S_0^\perp\) is bounded away from \(0\) along the trajectory, then it is manifest that \(\Delta\varphi_0 = \order{\vert\delta\vert}\). The trajectory almost forms a closed loop with period \(T\), up to a small error of order \(\vert\delta\vert\).
In the next period from \(T\) to \(2T\), the integrand of \autoref{eq:Delta_Phi} is also perturbed by \(\order{\delta}\), so that the change in \(\varphi_0\) over two periods is \(2\Delta\varphi_0 + \order{\delta^2}\). 
The accumulation of these changes over many cycles yields a linear in time drift of \(\varphi_0\) (and similarly of $\tau_0$) at an effective rate of $\vert\Delta\varphi_0/T\vert \propto \vert\delta\vert$.
One can apply a similar analysis to other integrals of motion. The conclusion about linear drift remains valid even if \(S_0^\perp\) becomes temporarily small while $\eta$ and hence $\Bperp$ are still large.  In this case, the central spin trajectory passes close to the poles but quickly moves past them, closely following a great circle with speed \(B^\perp\). As a result, the azimuthal angle \(\varphi_0\) jumps rapidly by $\pi$ without affecting the conserved quantity $\tau_0$\footnote{While \autoref{eq:Delta_Phi} and \autoref{eq:dot_tau_primitive} have small denominators when \(S_0^\perp \approx 0\), we note that the numerator is also small because \(\HCS\) (appearing in \autoref{eq:dot_tau_primitive}) is proportional to \(S_0^\perp\). This results in a partial cancellation, and a finite change of \(\Delta \varphi_0 \approx \pi\) when \(B^\perp\) is large, which does not change $\tau_0$.}.

Linear drifts are not present in conventional integrable systems. 
Systems which have \(L\) independent conserved quantities in involution, like the Ishimori chain, have invariant tori which are invariant under Hamiltonian flows generated by \emph{any} of the conserved quantities, and not just the actual Hamiltonian~\cite{kubo_statistical_1957,Arnold1989_classicalmechanics}.
Then, one can show that if \(\tau\) is conserved and \(V\) is a perturbation, then the integral of \(\poissonbracket{\tau}{V}\) over the torus vanishes.
Indeed, let \(\rho\) be an invariant distribution with respect to all the conserved quantities, so that \(\poissonbracket{\tau}{\rho}=0\). This condition is satisfied for any GGE ensemble. Then
\begin{equation}\label{eqn:IntegrableGGENoDrift}
    \int \poissonbracket{\tau}{V} \rho \, \dd\mathbf{S} 
    = \int (\poissonbracket{\tau}{V \rho} - V \poissonbracket{\tau}{\rho})\, \dd\mathbf{S}
    = 0,
\end{equation}
where we used \(\poissonbracket{\tau}{\rho}=0\), the Leibniz rule for the Poisson bracket, and the fact that the integral of a Poisson bracket over all of phase space is zero.
Thus, the linear drift in conventional integrable systems vanishes, and conserved quantities decay due to the diffusion of trajectories between invariant tori, which gives the \(\delta^{-2}\) scaling of the melting time. 

\paragraph*{$\rm (ii)$} The angular separation, $\alpha$ can be seen in \figref{fig:example}{c}. While the distance in the quasi-conserved quantities (yellow) remains small, the two trajectories mainly separate in the angular coordinate (red). We can not resolve the functional form of this separation.

\paragraph*{Chaotic manifold.} 
The effect of the perturbation on trajectories changes dramatically when the parameter $\eta$ becomes small. As $\abs{\Sperp}$ is small, the environment spins precess around magnetic fields of strength $\delta h_j$ pointing in the $z$-direction (see \autoref{eq:EOM_CS_Pert}). 
This dynamics destroys the synchronization of the environmental spins necessary for regular motion, until $\eta$ again becomes large and the system resumes its quasi-integrable dynamics. These qualitative considerations agree with numerical results suggesting that in this chaotic manifold there are (i) large changes in quasi-conserved quantities and (ii) rapid separation of nearby trajectories within the chaotic manifold as shown in \autoref{fig:example}. 

We extract several lessons from \autoref{fig:example}. First, we find that an encounter with the chaotic manifold typically occurs with frequency \(\order{\vert\delta\vert}\). In \figref{fig:example}{a}, this occurs near \(t \approx 2\vert\delta\vert^{-1}\), where $\eta$ approaches zero. 
\figref{fig:example}{b} shows that \(\tau_0\) indeed undergoes a rapid jump when \(\eta\approx 0\).
\figref{fig:example}{c} shows the separation between two initially nearby trajectories, which exhibit a rapid divergence again when \(\eta \approx 0\). To quantify the separation, we use both the Euclidean distance in spin coordinates (blue) and the difference of the corresponding integrals of motion (yellow). Notably, the distance in $\tau_0$ increases by several orders of magnitude when $\eta$ approaches zero. The separation of the angles $\alpha$ also rapidly increases (shown in red), on top of a slow, steady drift in the quasi-integrable regime. The acquired difference in $\tau_0$ leads to a faster separation in $\alpha$ after the encounter. 

To conclude, we return to \figref{fig:time_scales}{b}.
In the quasi-integrable regime, trajectories closely follow the $\delta = 0$ regular trajectories, giving rise to nearly-closed one-dimensional orbits that are traversed on timescales of order $J^{-1}$. The integrals of motion experience a slow linear drift with speed proportional to $\vert\delta\vert$, resulting in a helical motion: displacement along the orthogonal directions (spanned by the quasi-conserved quantities) proceeds at a rate proportional to $\delta$, while the circular motion persists at a characteristic frequency set by $J$.
Upon approaching the chaotic manifold, the system exhibits strong divergence of nearby trajectories in all phase-space directions and rapid changes in the integrals of motion.

\begin{figure}[t]
    \centering
    \includegraphics[width=\columnwidth]{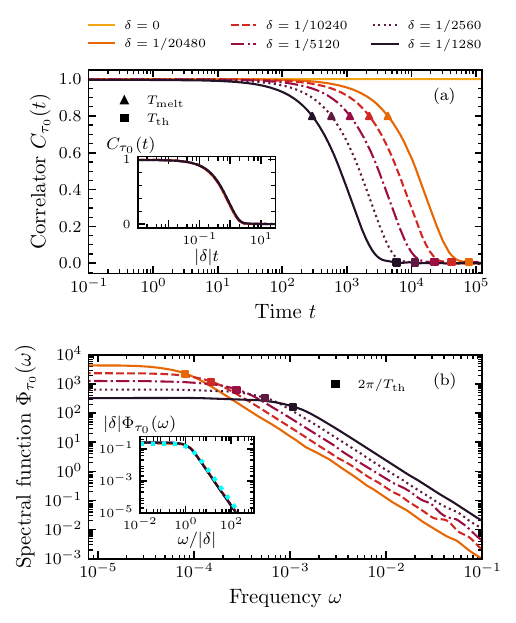}%
    \caption[]{Dynamics of $\tau_0$ as defined in \autoref{eq:centralspin_observable} in the central spin model with system size $L = 46$ at varying perturbation strengths $\delta$. Panels (a) and (b) show the autocorrelators and spectral functions of $\tau_0$, respectively. The triangles and squares mark the melting time $\Tmelt$ and thermalization time $\Tth$, respectively. In (a), the correlator is constant, $C_{\tau_0}(t) = 1$, at $\delta = 0$ (shown in light orange). The insets show rescaled data, indicating $\Tmelt, \Tth \propto \abs{\delta}^{-1}$. The inset in (b) further shows that the Lorentzian function (blue dotted line) is a good fit to $\Phi_{\tau_0}(\omega)$.
    \label{fig:centralspin_cspec}}
\end{figure}

\begin{figure}[t]
    \centering
    \includegraphics[width=\columnwidth]{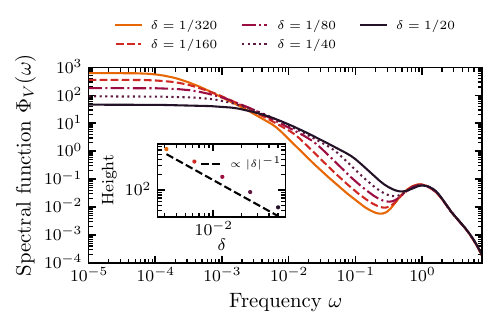}%
    \caption[]{Spectral function of $V$ with varying perturbation strengths for the central spin model (with $L = 46$). The profile of $\Phi_{V}(\omega)$ sharpens as $\delta$ decreases due to the presence of a nonzero Drude weight for $\delta = 0$. The inset shows the height of the spectral function at $\omega = 0$ for varying $\delta$, which scales as $1/\Gamma \propto \vert\delta\vert^{-1}$ (black dashed line).
    \label{fig:central_spin_spec_V}}
\end{figure} 

\subsection{Decay of autocorrelators}\label{ssec:decay_AC}
The perturbed central spin model at zero energy relaxes on a timescale inversely proportional to the perturbation strength, \( \Tmelt, \Tth \propto \vert\delta\vert^{-1} \). The decay of the autocorrelator of $O=\tau_0$ is illustrated in \autoref{fig:centralspin_cspec}, which shows excellent scaling collapses of both the autocorrelator and the spectral function in the insets. Similarly, other observables decay on a timescale proportional to $\vert\delta\vert^{-1}$, as shown in \aref{app:non_conserved_observables}.

The observed scaling is faster than anticipated; standard perturbative arguments predict $\Tmelt, \Tth \propto \delta^{-2}$, which holds for the Ishimori chain. In \aref{app:lower_bound}, we prove that for quantities which have a non-singular Poisson bracket with the Hamiltonian, \(\Tth \propto \vert\delta\vert^{-1}\) is the fastest possible scaling of the thermalization. We note, however, that this proof does not directly apply to $\tau_0$ because its equations of motion have singularities.

We can understand the breakdown of the perturbative treatment in \autoref{eq:FGR} through the existence of a nonzero Drude peak in the spectral function of the perturbation \(V\) at $\delta =0$:
\begin{equation}
    \Phi_{V}(\omega) = D_{V} \delta(\omega) + \Phi_{V}^{\mathrm{reg}}(\omega).\label{eq:delta_function}
\end{equation}
Note that $\delta(\omega)$ in \autoref{eq:delta_function} denotes a delta-function and not the perturbation strength.
The Drude peak in $\Phi_{V}(\omega)$ is shown in \autoref{fig:central_spin_spec_V}.

The Drude weight can be self-consistently regularized to obtain a finite prediction for the relaxation rate.
Suppose that the integrability-breaking perturbation broadens the Drude peak in \(\Phi_{V}(\omega)\) to a Lorentzian with a width \(\Gamma_V\).
Then we use this renormalized spectral function in a standard perturbative calculation to find the decay rate of $\tau_0$: \(\Gamma_{\tau_0} \propto \delta^2\Phi_{V}(0) \propto \delta^2 \Gamma_V^{-1}\). 
Assuming that the broadening of the Drude peaks associated with all observables is the same, i.e., assuming that the system thermalizes on a single timescale $1/\Gamma$ with \(\Gamma = \Gamma_{\tau_0}=\Gamma_V\), we must have 
\begin{equation}
    \Gamma\propto \delta^2 /\Gamma \quad \implies\quad \Gamma \propto \vert\delta\vert,
\end{equation}
coinciding with the scaling of \(\Tmelt^{-1}\) and \(\Tth^{-1}\) we observe. 
This more phenomenological argument reproduces the predicted scaling from the analysis of individual trajectories in the previous section.

Notice that superintegrability is very important in this argument, as in normal integrable models, the conserved part of $V$ commutes (has a vanishing Poisson bracket) with the stationary GGE ensemble and does not contribute to the decay of $\tau_0$, as in \autoref{eqn:IntegrableGGENoDrift}. The microscopic mechanism of decay, as explained in the previous section, is facilitated by encounters with the chaotic manifold characterized by rapid  changes in conserved quantities.

While the initial memory of the autocorrelator is essentially lost after a few encounters with the chaotic manifold, the time to relax to the microcanonical ensemble value (although also scaling with $\vert\delta\vert^{-1}$) is significantly longer, as illustrated in \aref{app:supp_CS}.

\subsection{Divergence of trajectories}\label{ssec:divergence}
\begin{figure}[t]
    \centering
    \includegraphics[width=\columnwidth]{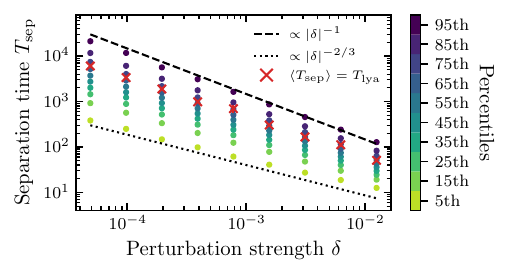}
	\caption{Separation time $\Tsep$ in the central spin model (with system size $L = 46$) at varying perturbation strength $\delta$. Different colors show different percentiles of the $\Tsep$ distribution across the phase space. $\Tlya$ is shown in red crosses. The lowest percentile scales approximately as $\vert\delta\vert^{-2/3}$ while the highest percentile scales as $\vert\delta\vert^{-1}$, as does the mean.\label{fig:separation}}
\end{figure}

The divergence of trajectories can also be attributed to encounters with the chaotic manifold. Because such encounters are rare, we observe a broad distribution of separation times \(\Tsep\) (defined in \autoref{eq:tsep}) as illustrated in \autoref{fig:separation}. 

Unlike in the Ishimori chain, a complete collapse of the distribution of separation times, as seen in \autoref{fig:ishimori_tlya}, is not possible. 
The mean (red crosses) corresponds to the Lyapunov times $\Tlya$ and agrees with the timescale associated with encountering the chaotic manifold, $\vert\delta\vert^{-1}$.

Within the quasi-integrable regime, nearby trajectories predominantly separate along the angular directions associated with the integrable dynamics (red curve in \figref{fig:example}{c}). This separation is slow. Nevertheless, it resembles confined separation—similar to what is observed in the Ishimori chain and in our solar system~\cite{milani_an_1992,milani_stable_1997,laskar_chaotic_2008,winter_short_2010,pereira_confined_2024}—in the sense that trajectories diverge within quasi-invariant tori, constrained by the presence of quasi-conserved quantities. 

Once the trajectory enters the chaotic manifold, the situation changes dramatically, and the separation between trajectories essentially jumps. This is illustrated in \figref{fig:example}{c} where the separation in $\tau_0$ jumps in the grey region from $10^{-10}$ to $10^{-6}$. Empirical observations suggest that the size of the jump is determined by the previously achieved separation in the angle variable. Because separation between trajectories occurs in a direction which is not constrained by any approximate conservation laws, there is no confinement of the two nearby trajectories to a particular manifold, as happens in the Ishimori chain. This justifies terming the associated phenomenology as \emph{deconfined chaos}.

\subsection{Deconfined Chaos and the SYK model}
\label{ssec:SYK}
The perturbed central spin model shares dynamical properties with certain quantum systems that exhibit fast thermalization, like the Sachdev-Ye-Kitaev (SYK) model, strange metals, or models of black holes~\cite{Sachdev_2024}. The analogy is seen by identifying temperature $T_{\rm temp}$ in these quantum models with the integrability-breaking parameter $\delta$ in the classical models we analyze here. 

In conventional systems like Fermi liquids, the relaxation rate (and hence resistivity) scales as $T_{\rm temp}^2/J$, where $J$ is some fast microscopic energy scale set, e.g., by the Fermi energy. Likewise, in the conventional confined chaos scenario realized, for example, in the Ishimori chain, we have $\Gamma\propto \delta^2/J$. This similarity is not a coincidence, as both results follow from perturbation theory (FGR or more generally, kinetic approaches). 

In the SYK model and other mentioned models, the relaxation rate is linear in temperature $\Gamma\propto T_{\rm temp}$ and does not depend on the microscopic scale $J$. Similarly, in the central spin model, the relaxation rate $\Gamma\propto \vert\delta\vert$ and also does not depend on $J$. 
Both in the SYK model and the central spin model, this rate saturates the upper bound for the relaxation rate, so both models are maximally ergodic. 

Further, in both the SYK model and the central spin model, the Lyapunov scale defining the separation time for initially nearby trajectories---as measured by the out-of-time-order correlator (OTOC) in SYK---also scales as $T_{\rm temp}$ or $\vert\delta\vert$, respectively. This is to be contrasted with the confined chaos models where the separation time is either much shorter than the relaxation time, like in the Ishimori chain, or is not even defined (as there is no exponential growth of the OTOC)~\cite{Fine_2014, Kukuljan_2017}. 

More studies are needed to determine whether these parallels go beyond mere observation. Classical systems, such as the ones studied here, provide a detailed microscopic mechanism for deconfined chaos. It remains to be seen whether the associated properties, e.g. very broad distributions of timescales, can be extended to quantum models. 
For the central spin model, the dynamics is expected to be the same (in the large $L$ limit) irrespective of the spin size~\cite{tang_integrability_2023}. 
Conversely, in Refs.~\cite {Davidson_2017,Schmitt_2019}, it was found that the dynamics of the SYK model is well described by the truncated Wigner approximation (TWA) at infinite temperatures, where time evolution is purely classical. Therefore, such an extension to quantum regimes should be possible, and the deconfined chaos mechanism should be applicable to quantum superintegrable models as well.

\section{Conclusion} \label{sec:conclusion}
We have described deconfined chaos, a new scenario for thermalization in weakly perturbed integrable systems (see \autoref{fig:time_scales}). 
Systems with deconfined chaos exhibit Lyapunov instabilities and relaxational dynamics on the same characteristic timescale, in contrast to confined chaotic systems in which Lyapunov instabilities asymptotically precede relaxation (see \autoref{fig:real_time_scales}).
We understand deconfined chaos to be a consequence of the inhomogeneity of phase space; upon weak perturbation, chaos is restricted to a thin manifold in which the natural timescales of the integrable dynamics are sufficiently long. As trajectories traverse this manifold, they are very sensitive to the perturbation, and few-body observables (and in particular conserved quantities associated with the integrable point) relax during the traversal. We have presented two case studies in classical spin systems, one each for the scenarios of confined and deconfined chaos. 

A secondary result is the unusual phase-space structure of the classical central spin model with XX interactions. It is maximally superintegrable in a single energy shell, and empirically non-integrable in all other energy shells. We are not aware of other models with this structure. 

A third result is that the perturbed central spin model relaxes on the fastest possible timescale, $\Tth\propto \vert\delta\vert^{-1}$, where $\delta$ is the strength of the perturbation. This is reminiscent of maximally ergodic models, such as the SYK model. We find that it is the superintegrability of the central spin model that underlies the emergence of this fast non-perturbative relaxation. Superintegrability is responsible for macroscopic degeneracies of unperturbed orbits (energy eigenstates in quantum language), leading to their linear drift under weak perturbations and rapid decay upon encountering the thin chaotic manifold. From the strong similarity between the relaxation of central spin and SYK models, we speculate that the zero temperature limit of the SYK model may also be superintegrable. Further supporting this speculation is the fact that the SYK model has a nonzero entropy density as zero temperature is approached, which is reminiscent of the frozen macroscopically degenerate manifold in configurations at zero energy in the central spin model.

Extension of the phenomenology of confined and deconfined chaos to quantum models is an interesting future direction. Instead of studying Lyapunov instabilities, which often do not show up in quantum systems~\cite{wijn_lyapunov_2013, Kukuljan_2017}, one can analyze two-point correlations between angle variables. The connections to SYK-like models and further consequences for quantum models need further exploration.

Another avenue for future work is the necessary and sufficient conditions for deconfined chaos. In our example, the essential ingredient is the inhomogeneity of phase space, which is a consequence of maximal superintegrability in an energy shell and the mean-field character of the model. Determining whether these properties are necessary for deconfined chaos in classical and quantum settings is an interesting open question.

Another direction is to sharpen how energy may be treated as an effective ``integrability-breaking perturbation'' in the unperturbed central spin model with XX interactions. At generic values of the energy, numerical simulations find features reminiscent of glassy dynamics: long thermalization times, a power-law decay of correlation functions, and inhomogeneous phase space portraits with chaotic and regular regions.
While the $\delta$-perturbation we analyzed here leads to maximally fast relaxation of the central spin model, perturbing the model differently leads to an opposite regime of  slow thermalization. This is reminiscent of ``sticky'' dynamics near the boundaries between chaotic and regular regions in phase space~\cite{meiss_markov-tree_1985,srivastava_integrable_1988}; the XX central spin model may provide us with a many-body example of stickiness.

\section*{Acknowledgement}
The authors thank Daniel Arovas, Pieter Claeys, Michael Flynn, Masud Haque, Jorge Kurchan, Gerhard M\"uller, and Joachim Stolze for helpful discussions, and Tongyu Zhou for creating the schematic illustrations. This work was supported by: NSF grant no. DMR-2412542 (HK, AP), AFOSR grant nos. FA9550-21-1-0342 (HK, AP) and FA9550-24-1-0121 (RS, AC), 
a Stanford Q-FARM Bloch Fellowship (DML), and a Packard Fellowship in Science and Engineering (DML, PI: Vedika Khemani).
Numerical simulations were carried out using the Julia packages \texttt{Sobol.jl}~\cite{Soboljl}, \texttt{ChaosTools.jl}~\cite{ChaosTools}, and \texttt{DynamicalSystems.jl}~\cite{Datseris2018,DatserisParlitz2022}.

\bibliography{ref.bib}

\begin{thebibliography}{72}%
\makeatletter
\providecommand \@ifxundefined [1]{%
 \@ifx{#1\undefined}
}%
\providecommand \@ifnum [1]{%
 \ifnum #1\expandafter \@firstoftwo
 \else \expandafter \@secondoftwo
 \fi
}%
\providecommand \@ifx [1]{%
 \ifx #1\expandafter \@firstoftwo
 \else \expandafter \@secondoftwo
 \fi
}%
\providecommand \natexlab [1]{#1}%
\providecommand \enquote  [1]{``#1''}%
\providecommand \bibnamefont  [1]{#1}%
\providecommand \bibfnamefont [1]{#1}%
\providecommand \citenamefont [1]{#1}%
\providecommand \href@noop [0]{\@secondoftwo}%
\providecommand \href [0]{\begingroup \@sanitize@url \@href}%
\providecommand \@href[1]{\@@startlink{#1}\@@href}%
\providecommand \@@href[1]{\endgroup#1\@@endlink}%
\providecommand \@sanitize@url [0]{\catcode `\\12\catcode `\$12\catcode `\&12\catcode `\#12\catcode `\^12\catcode `\_12\catcode `\%12\relax}%
\providecommand \@@startlink[1]{}%
\providecommand \@@endlink[0]{}%
\providecommand \url  [0]{\begingroup\@sanitize@url \@url }%
\providecommand \@url [1]{\endgroup\@href {#1}{\urlprefix }}%
\providecommand \urlprefix  [0]{URL }%
\providecommand \Eprint [0]{\href }%
\providecommand \doibase [0]{https://doi.org/}%
\providecommand \selectlanguage [0]{\@gobble}%
\providecommand \bibinfo  [0]{\@secondoftwo}%
\providecommand \bibfield  [0]{\@secondoftwo}%
\providecommand \translation [1]{[#1]}%
\providecommand \BibitemOpen [0]{}%
\providecommand \bibitemStop [0]{}%
\providecommand \bibitemNoStop [0]{.\EOS\space}%
\providecommand \EOS [0]{\spacefactor3000\relax}%
\providecommand \BibitemShut  [1]{\csname bibitem#1\endcsname}%
\let\auto@bib@innerbib\@empty
\bibitem [{\citenamefont {Landau}\ and\ \citenamefont {Lifshitz}(1976)}]{LANDAU:1976aa}%
  \BibitemOpen
  \bibfield  {author} {\bibinfo {author} {\bibfnamefont {L.}~\bibnamefont {Landau}}\ and\ \bibinfo {author} {\bibfnamefont {E.}~\bibnamefont {Lifshitz}},\ }\href@noop {} {\emph {\bibinfo {title} {Mechanics}}},\ \bibinfo {series} {Course of Theoretical Physics}, Vol.~\bibinfo {volume} {1}\ (\bibinfo  {publisher} {Butterworth-Heinemann},\ \bibinfo {year} {1976})\BibitemShut {NoStop}%
\bibitem [{\citenamefont {Kinoshita}\ \emph {et~al.}(2006)\citenamefont {Kinoshita}, \citenamefont {Wenger},\ and\ \citenamefont {Weiss}}]{kinoshita2006quantum}%
  \BibitemOpen
  \bibfield  {author} {\bibinfo {author} {\bibfnamefont {T.}~\bibnamefont {Kinoshita}}, \bibinfo {author} {\bibfnamefont {T.}~\bibnamefont {Wenger}},\ and\ \bibinfo {author} {\bibfnamefont {D.~S.}\ \bibnamefont {Weiss}},\ }\bibfield  {title} {\bibinfo {title} {A quantum newton's cradle},\ }\href {https://doi.org/10.1038/nature04693} {\bibfield  {journal} {\bibinfo  {journal} {Nature}\ }\textbf {\bibinfo {volume} {440}},\ \bibinfo {pages} {900} (\bibinfo {year} {2006})}\BibitemShut {NoStop}%
\bibitem [{\citenamefont {Tang}\ \emph {et~al.}(2018)\citenamefont {Tang}, \citenamefont {Kao}, \citenamefont {Li}, \citenamefont {Seo}, \citenamefont {Mallayya}, \citenamefont {Rigol}, \citenamefont {Gopalakrishnan},\ and\ \citenamefont {Lev}}]{tang2018thermalization}%
  \BibitemOpen
  \bibfield  {author} {\bibinfo {author} {\bibfnamefont {Y.}~\bibnamefont {Tang}}, \bibinfo {author} {\bibfnamefont {W.}~\bibnamefont {Kao}}, \bibinfo {author} {\bibfnamefont {K.-Y.}\ \bibnamefont {Li}}, \bibinfo {author} {\bibfnamefont {S.}~\bibnamefont {Seo}}, \bibinfo {author} {\bibfnamefont {K.}~\bibnamefont {Mallayya}}, \bibinfo {author} {\bibfnamefont {M.}~\bibnamefont {Rigol}}, \bibinfo {author} {\bibfnamefont {S.}~\bibnamefont {Gopalakrishnan}},\ and\ \bibinfo {author} {\bibfnamefont {B.~L.}\ \bibnamefont {Lev}},\ }\bibfield  {title} {\bibinfo {title} {Thermalization near integrability in a dipolar quantum newton's cradle},\ }\href {https://doi.org/10.1103/PhysRevX.8.021030} {\bibfield  {journal} {\bibinfo  {journal} {Physical Review X}\ }\textbf {\bibinfo {volume} {8}},\ \bibinfo {pages} {021030} (\bibinfo {year} {2018})}\BibitemShut {NoStop}%
\bibitem [{\citenamefont {Malvania}\ \emph {et~al.}(2021)\citenamefont {Malvania}, \citenamefont {Zhang}, \citenamefont {Le}, \citenamefont {Dubail}, \citenamefont {Rigol},\ and\ \citenamefont {Weiss}}]{Malvania:2021aa}%
  \BibitemOpen
  \bibfield  {author} {\bibinfo {author} {\bibfnamefont {N.}~\bibnamefont {Malvania}}, \bibinfo {author} {\bibfnamefont {Y.}~\bibnamefont {Zhang}}, \bibinfo {author} {\bibfnamefont {Y.}~\bibnamefont {Le}}, \bibinfo {author} {\bibfnamefont {J.}~\bibnamefont {Dubail}}, \bibinfo {author} {\bibfnamefont {M.}~\bibnamefont {Rigol}},\ and\ \bibinfo {author} {\bibfnamefont {D.~S.}\ \bibnamefont {Weiss}},\ }\bibfield  {title} {\bibinfo {title} {Generalized hydrodynamics in strongly interacting 1d bose gases},\ }\href {https://www.science.org/doi/abs/10.1126/science.abf0147} {\bibfield  {journal} {\bibinfo  {journal} {Science}\ }\textbf {\bibinfo {volume} {373}},\ \bibinfo {pages} {1129} (\bibinfo {year} {2021})}\BibitemShut {NoStop}%
\bibitem [{\citenamefont {Ott}(2002)}]{Ott_2002}%
  \BibitemOpen
  \bibfield  {author} {\bibinfo {author} {\bibfnamefont {E.}~\bibnamefont {Ott}},\ }\href {https://doi.org/10.1017/CBO9780511803260} {\emph {\bibinfo {title} {Chaos in Dynamical Systems}}},\ \bibinfo {edition} {2nd}\ ed.\ (\bibinfo  {publisher} {Cambridge University Press},\ \bibinfo {year} {2002})\BibitemShut {NoStop}%
\bibitem [{\citenamefont {Strogatz}(2018)}]{strogatz2018nonlinear}%
  \BibitemOpen
  \bibfield  {author} {\bibinfo {author} {\bibfnamefont {S.}~\bibnamefont {Strogatz}},\ }\href {https://doi.org/10.1201/9780429398490} {\emph {\bibinfo {title} {Nonlinear Dynamics and Chaos: With Applications to Physics, Biology, Chemistry, and Engineering}}}\ (\bibinfo  {publisher} {CRC Press},\ \bibinfo {year} {2018})\BibitemShut {NoStop}%
\bibitem [{\citenamefont {Kollar}\ \emph {et~al.}(2011)\citenamefont {Kollar}, \citenamefont {Wolf},\ and\ \citenamefont {Eckstein}}]{kollar_generalized_2011}%
  \BibitemOpen
  \bibfield  {author} {\bibinfo {author} {\bibfnamefont {M.}~\bibnamefont {Kollar}}, \bibinfo {author} {\bibfnamefont {F.~A.}\ \bibnamefont {Wolf}},\ and\ \bibinfo {author} {\bibfnamefont {M.}~\bibnamefont {Eckstein}},\ }\bibfield  {title} {\bibinfo {title} {Generalized gibbs ensemble prediction of prethermalization plateaus and their relation to nonthermal steady states in integrable systems},\ }\href {https://doi.org/10.1103/PhysRevB.84.054304} {\bibfield  {journal} {\bibinfo  {journal} {Phys. Rev. B}\ }\textbf {\bibinfo {volume} {84}},\ \bibinfo {pages} {054304} (\bibinfo {year} {2011})}\BibitemShut {NoStop}%
\bibitem [{\citenamefont {Vidmar}\ and\ \citenamefont {Rigol}(2016)}]{vidmar_generalized_2016}%
  \BibitemOpen
  \bibfield  {author} {\bibinfo {author} {\bibfnamefont {L.}~\bibnamefont {Vidmar}}\ and\ \bibinfo {author} {\bibfnamefont {M.}~\bibnamefont {Rigol}},\ }\bibfield  {title} {\bibinfo {title} {Generalized gibbs ensemble in integrable lattice models},\ }\href {https://doi.org/10.1088/1742-5468/2016/06/064007} {\bibfield  {journal} {\bibinfo  {journal} {Journal of Statistical Mechanics: Theory and Experiment}\ }\textbf {\bibinfo {volume} {2016}},\ \bibinfo {pages} {064007} (\bibinfo {year} {2016})}\BibitemShut {NoStop}%
\bibitem [{\citenamefont {Rigol}\ \emph {et~al.}(2007)\citenamefont {Rigol}, \citenamefont {Dunjko}, \citenamefont {Yurovsky},\ and\ \citenamefont {Olshanii}}]{rigol2007relaxation}%
  \BibitemOpen
  \bibfield  {author} {\bibinfo {author} {\bibfnamefont {M.}~\bibnamefont {Rigol}}, \bibinfo {author} {\bibfnamefont {V.}~\bibnamefont {Dunjko}}, \bibinfo {author} {\bibfnamefont {V.}~\bibnamefont {Yurovsky}},\ and\ \bibinfo {author} {\bibfnamefont {M.}~\bibnamefont {Olshanii}},\ }\bibfield  {title} {\bibinfo {title} {Relaxation in a completely integrable many-body quantum system: An ab initio study of the dynamics of the highly excited states of 1d lattice hard-core bosons},\ }\href {https://doi.org/10.1103/PhysRevLett.98.050405} {\bibfield  {journal} {\bibinfo  {journal} {Phys. Rev. Lett.}\ }\textbf {\bibinfo {volume} {98}},\ \bibinfo {pages} {050405} (\bibinfo {year} {2007})}\BibitemShut {NoStop}%
\bibitem [{\citenamefont {Milani}\ and\ \citenamefont {Nobili}(1992)}]{milani_an_1992}%
  \BibitemOpen
  \bibfield  {author} {\bibinfo {author} {\bibfnamefont {A.}~\bibnamefont {Milani}}\ and\ \bibinfo {author} {\bibfnamefont {A.~M.}\ \bibnamefont {Nobili}},\ }\bibfield  {title} {\bibinfo {title} {An example of stable chaos in the solar system},\ }\href {https://doi.org/10.1038/357569a0} {\bibfield  {journal} {\bibinfo  {journal} {Nature}\ }\textbf {\bibinfo {volume} {357}},\ \bibinfo {pages} {569} (\bibinfo {year} {1992})}\BibitemShut {NoStop}%
\bibitem [{\citenamefont {Milani}\ \emph {et~al.}(1997)\citenamefont {Milani}, \citenamefont {Nobili},\ and\ \citenamefont {Knežević}}]{milani_stable_1997}%
  \BibitemOpen
  \bibfield  {author} {\bibinfo {author} {\bibfnamefont {A.}~\bibnamefont {Milani}}, \bibinfo {author} {\bibfnamefont {A.~M.}\ \bibnamefont {Nobili}},\ and\ \bibinfo {author} {\bibfnamefont {Z.}~\bibnamefont {Knežević}},\ }\bibfield  {title} {\bibinfo {title} {Stable chaos in the asteroid belt},\ }\href {https://doi.org/https://doi.org/10.1006/icar.1996.5582} {\bibfield  {journal} {\bibinfo  {journal} {Icarus}\ }\textbf {\bibinfo {volume} {125}},\ \bibinfo {pages} {13} (\bibinfo {year} {1997})}\BibitemShut {NoStop}%
\bibitem [{\citenamefont {Laskar}(2008)}]{laskar_chaotic_2008}%
  \BibitemOpen
  \bibfield  {author} {\bibinfo {author} {\bibfnamefont {J.}~\bibnamefont {Laskar}},\ }\bibfield  {title} {\bibinfo {title} {Chaotic diffusion in the solar system},\ }\href {https://doi.org/https://doi.org/10.1016/j.icarus.2008.02.017} {\bibfield  {journal} {\bibinfo  {journal} {Icarus}\ }\textbf {\bibinfo {volume} {196}},\ \bibinfo {pages} {1} (\bibinfo {year} {2008})}\BibitemShut {NoStop}%
\bibitem [{\citenamefont {Winter}\ \emph {et~al.}(2010)\citenamefont {Winter}, \citenamefont {Mourão},\ and\ \citenamefont {Winter}}]{winter_short_2010}%
  \BibitemOpen
  \bibfield  {author} {\bibinfo {author} {\bibfnamefont {O.~C.}\ \bibnamefont {Winter}}, \bibinfo {author} {\bibfnamefont {D.~C.}\ \bibnamefont {Mourão}},\ and\ \bibinfo {author} {\bibfnamefont {S.~M.~G.}\ \bibnamefont {Winter}},\ }\bibfield  {title} {\bibinfo {title} {Short {Lyapunov} time: a method for identifying confined chaos},\ }\href {https://doi.org/10.1051/0004-6361/200912734} {\bibfield  {journal} {\bibinfo  {journal} {Astronomy \& Astrophysics}\ }\textbf {\bibinfo {volume} {523}},\ \bibinfo {pages} {A67} (\bibinfo {year} {2010})}\BibitemShut {NoStop}%
\bibitem [{\citenamefont {Pereira}\ \emph {et~al.}(2024)\citenamefont {Pereira}, \citenamefont {Mourão},\ and\ \citenamefont {Winter}}]{pereira_confined_2024}%
  \BibitemOpen
  \bibfield  {author} {\bibinfo {author} {\bibfnamefont {L.~S.}\ \bibnamefont {Pereira}}, \bibinfo {author} {\bibfnamefont {D.~C.}\ \bibnamefont {Mourão}},\ and\ \bibinfo {author} {\bibfnamefont {O.~C.}\ \bibnamefont {Winter}},\ }\bibfield  {title} {\bibinfo {title} {Confined chaos and the chaotic angular motion of atlas, a saturn’s inner satellite},\ }\href {https://doi.org/10.1093/mnras/stae457} {\bibfield  {journal} {\bibinfo  {journal} {Monthly Notices of the Royal Astronomical Society}\ }\textbf {\bibinfo {volume} {529}},\ \bibinfo {pages} {1012} (\bibinfo {year} {2024})}\BibitemShut {NoStop}%
\bibitem [{\citenamefont {Goldfriend}\ and\ \citenamefont {Kurchan}(2020)}]{goldfriend2020quasi}%
  \BibitemOpen
  \bibfield  {author} {\bibinfo {author} {\bibfnamefont {T.}~\bibnamefont {Goldfriend}}\ and\ \bibinfo {author} {\bibfnamefont {J.}~\bibnamefont {Kurchan}},\ }\bibfield  {title} {\bibinfo {title} {Quasi-integrable systems are slow to thermalize but may be good scramblers},\ }\href {https://doi.org/10.1103/PhysRevE.102.022201} {\bibfield  {journal} {\bibinfo  {journal} {Phys. Rev. E}\ }\textbf {\bibinfo {volume} {102}},\ \bibinfo {pages} {022201} (\bibinfo {year} {2020})}\BibitemShut {NoStop}%
\bibitem [{\citenamefont {Benettin}\ \emph {et~al.}(2013)\citenamefont {Benettin}, \citenamefont {Christodoulidi},\ and\ \citenamefont {Ponno}}]{benettin_the_2013}%
  \BibitemOpen
  \bibfield  {author} {\bibinfo {author} {\bibfnamefont {G.}~\bibnamefont {Benettin}}, \bibinfo {author} {\bibfnamefont {H.}~\bibnamefont {Christodoulidi}},\ and\ \bibinfo {author} {\bibfnamefont {A.}~\bibnamefont {Ponno}},\ }\bibfield  {title} {\bibinfo {title} {{The Fermi-Pasta-Ulam Problem and Its Underlying Integrable Dynamics}},\ }\href {https://doi.org/10.1007/s10955-013-0760-6} {\bibfield  {journal} {\bibinfo  {journal} {Journal of Statistical Physics}\ }\textbf {\bibinfo {volume} {152}},\ \bibinfo {pages} {195} (\bibinfo {year} {2013})}\BibitemShut {NoStop}%
\bibitem [{\citenamefont {Benettin}\ \emph {et~al.}(2018)\citenamefont {Benettin}, \citenamefont {Pasquali},\ and\ \citenamefont {Ponno}}]{benettin_fermi_2018}%
  \BibitemOpen
  \bibfield  {author} {\bibinfo {author} {\bibfnamefont {G.}~\bibnamefont {Benettin}}, \bibinfo {author} {\bibfnamefont {S.}~\bibnamefont {Pasquali}},\ and\ \bibinfo {author} {\bibfnamefont {A.}~\bibnamefont {Ponno}},\ }\bibfield  {title} {\bibinfo {title} {{The Fermi--Pasta--Ulam Problem and Its Underlying Integrable Dynamics: An Approach Through Lyapunov Exponents}},\ }\href {https://doi.org/10.1007/s10955-018-2017-x} {\bibfield  {journal} {\bibinfo  {journal} {Journal of Statistical Physics}\ }\textbf {\bibinfo {volume} {171}},\ \bibinfo {pages} {521} (\bibinfo {year} {2018})}\BibitemShut {NoStop}%
\bibitem [{\citenamefont {Goldfriend}\ and\ \citenamefont {Kurchan}(2019)}]{goldfriend2019equilibration}%
  \BibitemOpen
  \bibfield  {author} {\bibinfo {author} {\bibfnamefont {T.}~\bibnamefont {Goldfriend}}\ and\ \bibinfo {author} {\bibfnamefont {J.}~\bibnamefont {Kurchan}},\ }\bibfield  {title} {\bibinfo {title} {Equilibration of quasi-integrable systems},\ }\href {https://doi.org/10.1103/PhysRevE.99.022146} {\bibfield  {journal} {\bibinfo  {journal} {Phys. Rev. E}\ }\textbf {\bibinfo {volume} {99}},\ \bibinfo {pages} {022146} (\bibinfo {year} {2019})}\BibitemShut {NoStop}%
\bibitem [{\citenamefont {Mishchenko}\ and\ \citenamefont {Fomenko}(1978)}]{Mishchenko1978superintegrable}%
  \BibitemOpen
  \bibfield  {author} {\bibinfo {author} {\bibfnamefont {A.~S.}\ \bibnamefont {Mishchenko}}\ and\ \bibinfo {author} {\bibfnamefont {A.~T.}\ \bibnamefont {Fomenko}},\ }\bibfield  {title} {\bibinfo {title} {{Generalized Liouville method of integration of Hamiltonian systems}},\ }\href {https://doi.org/10.1007/BF01076254} {\bibfield  {journal} {\bibinfo  {journal} {Functional Analysis and Its Applications}\ }\textbf {\bibinfo {volume} {12}},\ \bibinfo {pages} {113} (\bibinfo {year} {1978})}\BibitemShut {NoStop}%
\bibitem [{\citenamefont {Bolsinov}\ and\ \citenamefont {Jovanovi{\'{c}}}(2003)}]{Bolsinov2003noncommutative}%
  \BibitemOpen
  \bibfield  {author} {\bibinfo {author} {\bibfnamefont {A.~V.}\ \bibnamefont {Bolsinov}}\ and\ \bibinfo {author} {\bibfnamefont {B.}~\bibnamefont {Jovanovi{\'{c}}}},\ }\bibfield  {title} {\bibinfo {title} {{Noncommutative Integrability, Moment Map and Geodesic Flows}},\ }\href {https://doi.org/10.1023/A:1023023300665} {\bibfield  {journal} {\bibinfo  {journal} {Annals of Global Analysis and Geometry}\ }\textbf {\bibinfo {volume} {23}},\ \bibinfo {pages} {305} (\bibinfo {year} {2003})}\BibitemShut {NoStop}%
\bibitem [{\citenamefont {Miller}\ \emph {et~al.}(2013)\citenamefont {Miller}, \citenamefont {Post},\ and\ \citenamefont {Winternitz}}]{miller_classical_2013}%
  \BibitemOpen
  \bibfield  {author} {\bibinfo {author} {\bibfnamefont {W.}~\bibnamefont {Miller}}, \bibinfo {author} {\bibfnamefont {S.}~\bibnamefont {Post}},\ and\ \bibinfo {author} {\bibfnamefont {P.}~\bibnamefont {Winternitz}},\ }\bibfield  {title} {\bibinfo {title} {Classical and quantum superintegrability with applications},\ }\href {https://doi.org/10.1088/1751-8113/46/42/423001} {\bibfield  {journal} {\bibinfo  {journal} {Journal of Physics A: Mathematical and Theoretical}\ }\textbf {\bibinfo {volume} {46}},\ \bibinfo {pages} {423001} (\bibinfo {year} {2013})}\BibitemShut {NoStop}%
\bibitem [{\citenamefont {Ishimori}(1982)}]{ishimori_an_1982}%
  \BibitemOpen
  \bibfield  {author} {\bibinfo {author} {\bibfnamefont {Y.}~\bibnamefont {Ishimori}},\ }\bibfield  {title} {\bibinfo {title} {An integrable classical spin chain},\ }\href {https://doi.org/10.1143/JPSJ.51.3417} {\bibfield  {journal} {\bibinfo  {journal} {Journal of the Physical Society of Japan}\ }\textbf {\bibinfo {volume} {51}},\ \bibinfo {pages} {3417} (\bibinfo {year} {1982})}\BibitemShut {NoStop}%
\bibitem [{\citenamefont {Theodorakopoulos}(1995)}]{Theodorakopoulos_nontopological_1995}%
  \BibitemOpen
  \bibfield  {author} {\bibinfo {author} {\bibfnamefont {N.}~\bibnamefont {Theodorakopoulos}},\ }\bibfield  {title} {\bibinfo {title} {Nontopological thermal solitons in isotropic ferromagnetic lattices},\ }\href {https://doi.org/10.1103/PhysRevB.52.9507} {\bibfield  {journal} {\bibinfo  {journal} {Phys. Rev. B}\ }\textbf {\bibinfo {volume} {52}},\ \bibinfo {pages} {9507} (\bibinfo {year} {1995})}\BibitemShut {NoStop}%
\bibitem [{\citenamefont {McRoberts}\ \emph {et~al.}(2022)\citenamefont {McRoberts}, \citenamefont {Bilitewski}, \citenamefont {Haque},\ and\ \citenamefont {Moessner}}]{McRoberts_long_2022}%
  \BibitemOpen
  \bibfield  {author} {\bibinfo {author} {\bibfnamefont {A.~J.}\ \bibnamefont {McRoberts}}, \bibinfo {author} {\bibfnamefont {T.}~\bibnamefont {Bilitewski}}, \bibinfo {author} {\bibfnamefont {M.}~\bibnamefont {Haque}},\ and\ \bibinfo {author} {\bibfnamefont {R.}~\bibnamefont {Moessner}},\ }\bibfield  {title} {\bibinfo {title} {Long-lived solitons and their signatures in the classical heisenberg chain},\ }\href {https://doi.org/10.1103/PhysRevE.106.L062202} {\bibfield  {journal} {\bibinfo  {journal} {Phys. Rev. E}\ }\textbf {\bibinfo {volume} {106}},\ \bibinfo {pages} {L062202} (\bibinfo {year} {2022})}\BibitemShut {NoStop}%
\bibitem [{\citenamefont {Villazon}\ \emph {et~al.}(2020{\natexlab{a}})\citenamefont {Villazon}, \citenamefont {Chandran},\ and\ \citenamefont {Claeys}}]{villazon_integrability_2020}%
  \BibitemOpen
  \bibfield  {author} {\bibinfo {author} {\bibfnamefont {T.}~\bibnamefont {Villazon}}, \bibinfo {author} {\bibfnamefont {A.}~\bibnamefont {Chandran}},\ and\ \bibinfo {author} {\bibfnamefont {P.~W.}\ \bibnamefont {Claeys}},\ }\bibfield  {title} {\bibinfo {title} {Integrability and dark states in an anisotropic central spin model},\ }\href {https://doi.org/10.1103/PhysRevResearch.2.032052} {\bibfield  {journal} {\bibinfo  {journal} {Phys. Rev. Res.}\ }\textbf {\bibinfo {volume} {2}},\ \bibinfo {pages} {032052} (\bibinfo {year} {2020}{\natexlab{a}})}\BibitemShut {NoStop}%
\bibitem [{\citenamefont {Tang}\ \emph {et~al.}(2023)\citenamefont {Tang}, \citenamefont {Long}, \citenamefont {Polkovnikov}, \citenamefont {Chandran},\ and\ \citenamefont {Claeys}}]{tang_integrability_2023}%
  \BibitemOpen
  \bibfield  {author} {\bibinfo {author} {\bibfnamefont {L.~H.}\ \bibnamefont {Tang}}, \bibinfo {author} {\bibfnamefont {D.~M.}\ \bibnamefont {Long}}, \bibinfo {author} {\bibfnamefont {A.}~\bibnamefont {Polkovnikov}}, \bibinfo {author} {\bibfnamefont {A.}~\bibnamefont {Chandran}},\ and\ \bibinfo {author} {\bibfnamefont {P.~W.}\ \bibnamefont {Claeys}},\ }\bibfield  {title} {\bibinfo {title} {Integrability and quench dynamics in the spin-1 central spin {XX} model},\ }\href {https://doi.org/10.21468/SciPostPhys.15.1.030} {\bibfield  {journal} {\bibinfo  {journal} {SciPost Physics}\ }\textbf {\bibinfo {volume} {15}},\ \bibinfo {pages} {030} (\bibinfo {year} {2023})}\BibitemShut {NoStop}%
\bibitem [{\citenamefont {Villazon}\ \emph {et~al.}(2020{\natexlab{b}})\citenamefont {Villazon}, \citenamefont {Claeys}, \citenamefont {Pandey}, \citenamefont {Polkovnikov},\ and\ \citenamefont {Chandran}}]{villazon_persistent_2020}%
  \BibitemOpen
  \bibfield  {author} {\bibinfo {author} {\bibfnamefont {T.}~\bibnamefont {Villazon}}, \bibinfo {author} {\bibfnamefont {P.~W.}\ \bibnamefont {Claeys}}, \bibinfo {author} {\bibfnamefont {M.}~\bibnamefont {Pandey}}, \bibinfo {author} {\bibfnamefont {A.}~\bibnamefont {Polkovnikov}},\ and\ \bibinfo {author} {\bibfnamefont {A.}~\bibnamefont {Chandran}},\ }\bibfield  {title} {\bibinfo {title} {Persistent dark states in anisotropic central spin models},\ }\href {https://doi.org/10.1038/s41598-020-73015-1} {\bibfield  {journal} {\bibinfo  {journal} {Scientific Reports}\ }\textbf {\bibinfo {volume} {10}},\ \bibinfo {pages} {16080} (\bibinfo {year} {2020}{\natexlab{b}})}\BibitemShut {NoStop}%
\bibitem [{\citenamefont {Dimo}\ and\ \citenamefont {Faribault}(2022)}]{dimo_strong_2022}%
  \BibitemOpen
  \bibfield  {author} {\bibinfo {author} {\bibfnamefont {C.}~\bibnamefont {Dimo}}\ and\ \bibinfo {author} {\bibfnamefont {A.}~\bibnamefont {Faribault}},\ }\bibfield  {title} {\bibinfo {title} {Strong-coupling emergence of dark states in xx central spin models},\ }\href {https://doi.org/10.1103/PhysRevB.105.L121404} {\bibfield  {journal} {\bibinfo  {journal} {Phys. Rev. B}\ }\textbf {\bibinfo {volume} {105}},\ \bibinfo {pages} {L121404} (\bibinfo {year} {2022})}\BibitemShut {NoStop}%
\bibitem [{\citenamefont {Sachdev}(2024)}]{Sachdev_2024}%
  \BibitemOpen
  \bibfield  {author} {\bibinfo {author} {\bibfnamefont {S.}~\bibnamefont {Sachdev}},\ }\bibfield  {title} {\bibinfo {title} {Quantum statistical mechanics of the sachdev-ye-kitaev model and charged black holes},\ }\bibfield  {journal} {\bibinfo  {journal} {International Journal of Modern Physics B}\ }\textbf {\bibinfo {volume} {38}},\ \href {https://doi.org/10.1142/s0217979224300032} {10.1142/s0217979224300032} (\bibinfo {year} {2024})\BibitemShut {NoStop}%
\bibitem [{\citenamefont {Faddeev}\ and\ \citenamefont {Takhtajan}(1987)}]{faddeevHamiltonianMethodsTheory1987}%
  \BibitemOpen
  \bibfield  {author} {\bibinfo {author} {\bibfnamefont {L.~D.}\ \bibnamefont {Faddeev}}\ and\ \bibinfo {author} {\bibfnamefont {L.~A.}\ \bibnamefont {Takhtajan}},\ }\href {https://doi.org/10.1007/978-3-540-69969-9} {\emph {\bibinfo {title} {Hamiltonian {{Methods}} in the {{Theory}} of {{Solitons}}}}}\ (\bibinfo  {publisher} {Springer},\ \bibinfo {address} {Berlin, Heidelberg},\ \bibinfo {year} {1987})\BibitemShut {NoStop}%
\bibitem [{\citenamefont {Prosen}\ and\ \citenamefont {\ifmmode \check{Z}\else \v{Z}\fi{}unkovi\ifmmode~\check{c}\else \v{c}\fi{}}(2013)}]{prosen_macroscopic_2013}%
  \BibitemOpen
  \bibfield  {author} {\bibinfo {author} {\bibfnamefont {T.}~\bibnamefont {Prosen}}\ and\ \bibinfo {author} {\bibfnamefont {B.}~\bibnamefont {\ifmmode \check{Z}\else \v{Z}\fi{}unkovi\ifmmode~\check{c}\else \v{c}\fi{}}},\ }\bibfield  {title} {\bibinfo {title} {Macroscopic diffusive transport in a microscopically integrable hamiltonian system},\ }\href {https://doi.org/10.1103/PhysRevLett.111.040602} {\bibfield  {journal} {\bibinfo  {journal} {Phys. Rev. Lett.}\ }\textbf {\bibinfo {volume} {111}},\ \bibinfo {pages} {040602} (\bibinfo {year} {2013})}\BibitemShut {NoStop}%
\bibitem [{\citenamefont {Lim}\ \emph {et~al.}(2025)\citenamefont {Lim}, \citenamefont {Matirko}, \citenamefont {Kim}, \citenamefont {Polkovnikov},\ and\ \citenamefont {Flynn}}]{lim2024defining}%
  \BibitemOpen
  \bibfield  {author} {\bibinfo {author} {\bibfnamefont {C.}~\bibnamefont {Lim}}, \bibinfo {author} {\bibfnamefont {K.}~\bibnamefont {Matirko}}, \bibinfo {author} {\bibfnamefont {H.}~\bibnamefont {Kim}}, \bibinfo {author} {\bibfnamefont {A.}~\bibnamefont {Polkovnikov}},\ and\ \bibinfo {author} {\bibfnamefont {M.~O.}\ \bibnamefont {Flynn}},\ }\href {https://doi.org/10.48550/arXiv.2401.01927} {\bibinfo {title} {Defining classical and quantum chaos through adiabatic transformations}} (\bibinfo {year} {2025}),\ \Eprint {https://arxiv.org/abs/2401.01927} {arXiv:2401.01927} \BibitemShut {NoStop}%
\bibitem [{\citenamefont {Langen}\ \emph {et~al.}(2016)\citenamefont {Langen}, \citenamefont {Gasenzer},\ and\ \citenamefont {Schmiedmayer}}]{langen2016prethermalization}%
  \BibitemOpen
  \bibfield  {author} {\bibinfo {author} {\bibfnamefont {T.}~\bibnamefont {Langen}}, \bibinfo {author} {\bibfnamefont {T.}~\bibnamefont {Gasenzer}},\ and\ \bibinfo {author} {\bibfnamefont {J.}~\bibnamefont {Schmiedmayer}},\ }\bibfield  {title} {\bibinfo {title} {Prethermalization and universal dynamics in near-integrable quantum systems},\ }\href {https://doi.org/10.1088/1742-5468/2016/06/064009} {\bibfield  {journal} {\bibinfo  {journal} {Journal of Statistical Mechanics: Theory and Experiment}\ }\textbf {\bibinfo {volume} {2016}},\ \bibinfo {pages} {064009} (\bibinfo {year} {2016})}\BibitemShut {NoStop}%
\bibitem [{\citenamefont {Mori}\ \emph {et~al.}(2018)\citenamefont {Mori}, \citenamefont {Ikeda}, \citenamefont {Kaminishi},\ and\ \citenamefont {Ueda}}]{mori2018thermalization}%
  \BibitemOpen
  \bibfield  {author} {\bibinfo {author} {\bibfnamefont {T.}~\bibnamefont {Mori}}, \bibinfo {author} {\bibfnamefont {T.~N.}\ \bibnamefont {Ikeda}}, \bibinfo {author} {\bibfnamefont {E.}~\bibnamefont {Kaminishi}},\ and\ \bibinfo {author} {\bibfnamefont {M.}~\bibnamefont {Ueda}},\ }\bibfield  {title} {\bibinfo {title} {Thermalization and prethermalization in isolated quantum systems: A theoretical overview},\ }\href {https://doi.org/10.1088/1361-6455/aabcdf} {\bibfield  {journal} {\bibinfo  {journal} {Journal of Physics B: Atomic, Molecular and Optical Physics}\ }\textbf {\bibinfo {volume} {51}},\ \bibinfo {pages} {112001} (\bibinfo {year} {2018})}\BibitemShut {NoStop}%
\bibitem [{\citenamefont {Skokos}(2010)}]{skokos_the_2010}%
  \BibitemOpen
  \bibfield  {author} {\bibinfo {author} {\bibfnamefont {C.}~\bibnamefont {Skokos}},\ }\bibinfo {title} {The lyapunov characteristic exponents and their computation},\ in\ \href {https://doi.org/10.1007/978-3-642-04458-8_2} {\emph {\bibinfo {booktitle} {Dynamics of Small Solar System Bodies and Exoplanets}}},\ \bibinfo {editor} {edited by\ \bibinfo {editor} {\bibfnamefont {J.~J.}\ \bibnamefont {Souchay}}\ and\ \bibinfo {editor} {\bibfnamefont {R.}~\bibnamefont {Dvorak}}}\ (\bibinfo  {publisher} {Springer Berlin Heidelberg},\ \bibinfo {address} {Berlin, Heidelberg},\ \bibinfo {year} {2010})\ pp.\ \bibinfo {pages} {63--135}\BibitemShut {NoStop}%
\bibitem [{\citenamefont {Benettin}\ \emph {et~al.}(1980)\citenamefont {Benettin}, \citenamefont {Galgani}, \citenamefont {Giorgilli},\ and\ \citenamefont {Strelcyn}}]{benettin1980}%
  \BibitemOpen
  \bibfield  {author} {\bibinfo {author} {\bibfnamefont {G.}~\bibnamefont {Benettin}}, \bibinfo {author} {\bibfnamefont {L.}~\bibnamefont {Galgani}}, \bibinfo {author} {\bibfnamefont {A.}~\bibnamefont {Giorgilli}},\ and\ \bibinfo {author} {\bibfnamefont {J.-M.}\ \bibnamefont {Strelcyn}},\ }\bibfield  {title} {\bibinfo {title} {Lyapunov characteristic exponents for smooth dynamical systems and for hamiltonian systems; a method for computing all of them. part 1: Theory},\ }\href {https://doi.org/10.1007/BF02128236} {\bibfield  {journal} {\bibinfo  {journal} {Meccanica}\ }\textbf {\bibinfo {volume} {15}},\ \bibinfo {pages} {9} (\bibinfo {year} {1980})}\BibitemShut {NoStop}%
\bibitem [{\citenamefont {Benettin}\ \emph {et~al.}(1976)\citenamefont {Benettin}, \citenamefont {Galgani},\ and\ \citenamefont {Strelcyn}}]{benettin1976kolmogorov}%
  \BibitemOpen
  \bibfield  {author} {\bibinfo {author} {\bibfnamefont {G.}~\bibnamefont {Benettin}}, \bibinfo {author} {\bibfnamefont {L.}~\bibnamefont {Galgani}},\ and\ \bibinfo {author} {\bibfnamefont {J.-M.}\ \bibnamefont {Strelcyn}},\ }\bibfield  {title} {\bibinfo {title} {Kolmogorov entropy and numerical experiments},\ }\href {https://doi.org/10.1103/PhysRevA.14.2338} {\bibfield  {journal} {\bibinfo  {journal} {Phys. Rev. A}\ }\textbf {\bibinfo {volume} {14}},\ \bibinfo {pages} {2338} (\bibinfo {year} {1976})}\BibitemShut {NoStop}%
\bibitem [{\citenamefont {Larkin}\ and\ \citenamefont {Ovchinnikov}(1968)}]{Larkin:1968aa}%
  \BibitemOpen
  \bibfield  {author} {\bibinfo {author} {\bibfnamefont {A.}~\bibnamefont {Larkin}}\ and\ \bibinfo {author} {\bibfnamefont {Y.~N.}\ \bibnamefont {Ovchinnikov}},\ }\bibfield  {title} {\bibinfo {title} {Quasiclassical method in the theory of superconductivity},\ }\href@noop {} {\bibfield  {journal} {\bibinfo  {journal} {Zh. Eksp. Teor. Fiz.}\ }\textbf {\bibinfo {volume} {55}},\ \bibinfo {pages} {2262} (\bibinfo {year} {1968})}\BibitemShut {NoStop}%
\bibitem [{\citenamefont {Rozenbaum}\ \emph {et~al.}(2017)\citenamefont {Rozenbaum}, \citenamefont {Ganeshan},\ and\ \citenamefont {Galitski}}]{rozenbaum_lyapunov_2017}%
  \BibitemOpen
  \bibfield  {author} {\bibinfo {author} {\bibfnamefont {E.~B.}\ \bibnamefont {Rozenbaum}}, \bibinfo {author} {\bibfnamefont {S.}~\bibnamefont {Ganeshan}},\ and\ \bibinfo {author} {\bibfnamefont {V.}~\bibnamefont {Galitski}},\ }\bibfield  {title} {\bibinfo {title} {Lyapunov exponent and out-of-time-ordered correlator's growth rate in a chaotic system},\ }\href {https://doi.org/10.1103/PhysRevLett.118.086801} {\bibfield  {journal} {\bibinfo  {journal} {Phys. Rev. Lett.}\ }\textbf {\bibinfo {volume} {118}},\ \bibinfo {pages} {086801} (\bibinfo {year} {2017})}\BibitemShut {NoStop}%
\bibitem [{\citenamefont {Lam}\ and\ \citenamefont {Kurchan}(2014)}]{lam2014stochastic}%
  \BibitemOpen
  \bibfield  {author} {\bibinfo {author} {\bibfnamefont {K.-D. N.~T.}\ \bibnamefont {Lam}}\ and\ \bibinfo {author} {\bibfnamefont {J.}~\bibnamefont {Kurchan}},\ }\bibfield  {title} {\bibinfo {title} {Stochastic perturbation of integrable systems: a window to weakly chaotic systems},\ }\href {https://doi.org/10.1007/s10955-014-1030-y} {\bibfield  {journal} {\bibinfo  {journal} {Journal of Statistical Physics}\ }\textbf {\bibinfo {volume} {156}},\ \bibinfo {pages} {619} (\bibinfo {year} {2014})}\BibitemShut {NoStop}%
\bibitem [{\citenamefont {Falcioni}\ \emph {et~al.}(1991)\citenamefont {Falcioni}, \citenamefont {Marconi},\ and\ \citenamefont {Vulpiani}}]{falcioni_ergodic_1991}%
  \BibitemOpen
  \bibfield  {author} {\bibinfo {author} {\bibfnamefont {M.}~\bibnamefont {Falcioni}}, \bibinfo {author} {\bibfnamefont {U.~M.~B.}\ \bibnamefont {Marconi}},\ and\ \bibinfo {author} {\bibfnamefont {A.}~\bibnamefont {Vulpiani}},\ }\bibfield  {title} {\bibinfo {title} {Ergodic properties of high-dimensional symplectic maps},\ }\href {https://doi.org/10.1103/PhysRevA.44.2263} {\bibfield  {journal} {\bibinfo  {journal} {Phys. Rev. A}\ }\textbf {\bibinfo {volume} {44}},\ \bibinfo {pages} {2263} (\bibinfo {year} {1991})}\BibitemShut {NoStop}%
\bibitem [{\citenamefont {Mulansky}\ \emph {et~al.}(2011)\citenamefont {Mulansky}, \citenamefont {Ahnert}, \citenamefont {Pikovsky},\ and\ \citenamefont {Shepelyansky}}]{mulansky_strong_2011}%
  \BibitemOpen
  \bibfield  {author} {\bibinfo {author} {\bibfnamefont {M.}~\bibnamefont {Mulansky}}, \bibinfo {author} {\bibfnamefont {K.}~\bibnamefont {Ahnert}}, \bibinfo {author} {\bibfnamefont {A.}~\bibnamefont {Pikovsky}},\ and\ \bibinfo {author} {\bibfnamefont {D.~L.}\ \bibnamefont {Shepelyansky}},\ }\bibfield  {title} {\bibinfo {title} {Strong and weak chaos in weakly nonintegrable many-body hamiltonian systems},\ }\href {https://doi.org/10.1007/s10955-011-0335-3} {\bibfield  {journal} {\bibinfo  {journal} {Journal of Statistical Physics}\ }\textbf {\bibinfo {volume} {145}},\ \bibinfo {pages} {1256} (\bibinfo {year} {2011})}\BibitemShut {NoStop}%
\bibitem [{\citenamefont {Malishava}\ and\ \citenamefont {Flach}(2022)}]{merab_lyapunov_2022}%
  \BibitemOpen
  \bibfield  {author} {\bibinfo {author} {\bibfnamefont {M.}~\bibnamefont {Malishava}}\ and\ \bibinfo {author} {\bibfnamefont {S.}~\bibnamefont {Flach}},\ }\bibfield  {title} {\bibinfo {title} {Lyapunov spectrum scaling for classical many-body dynamics close to integrability},\ }\href {https://doi.org/10.1103/PhysRevLett.128.134102} {\bibfield  {journal} {\bibinfo  {journal} {Phys. Rev. Lett.}\ }\textbf {\bibinfo {volume} {128}},\ \bibinfo {pages} {134102} (\bibinfo {year} {2022})}\BibitemShut {NoStop}%
\bibitem [{\citenamefont {de~Wijn}\ \emph {et~al.}(2012)\citenamefont {de~Wijn}, \citenamefont {Hess},\ and\ \citenamefont {Fine}}]{wijn_largest_2012}%
  \BibitemOpen
  \bibfield  {author} {\bibinfo {author} {\bibfnamefont {A.~S.}\ \bibnamefont {de~Wijn}}, \bibinfo {author} {\bibfnamefont {B.}~\bibnamefont {Hess}},\ and\ \bibinfo {author} {\bibfnamefont {B.~V.}\ \bibnamefont {Fine}},\ }\bibfield  {title} {\bibinfo {title} {Largest lyapunov exponents for lattices of interacting classical spins},\ }\href {https://doi.org/10.1103/PhysRevLett.109.034101} {\bibfield  {journal} {\bibinfo  {journal} {Phys. Rev. Lett.}\ }\textbf {\bibinfo {volume} {109}},\ \bibinfo {pages} {034101} (\bibinfo {year} {2012})}\BibitemShut {NoStop}%
\bibitem [{\citenamefont {Ljubotina}\ \emph {et~al.}(2017)\citenamefont {Ljubotina}, \citenamefont {{\v{Z}}nidari{\v{c}}},\ and\ \citenamefont {Prosen}}]{ljubotina2017spin}%
  \BibitemOpen
  \bibfield  {author} {\bibinfo {author} {\bibfnamefont {M.}~\bibnamefont {Ljubotina}}, \bibinfo {author} {\bibfnamefont {M.}~\bibnamefont {{\v{Z}}nidari{\v{c}}}},\ and\ \bibinfo {author} {\bibfnamefont {T.}~\bibnamefont {Prosen}},\ }\bibfield  {title} {\bibinfo {title} {Spin diffusion from an inhomogeneous quench in an integrable system},\ }\href {https://doi.org/10.1038/ncomms16117} {\bibfield  {journal} {\bibinfo  {journal} {Nature communications}\ }\textbf {\bibinfo {volume} {8}},\ \bibinfo {pages} {16117} (\bibinfo {year} {2017})}\BibitemShut {NoStop}%
\bibitem [{\citenamefont {De~Nardis}\ \emph {et~al.}(2021)\citenamefont {De~Nardis}, \citenamefont {Gopalakrishnan}, \citenamefont {Vasseur},\ and\ \citenamefont {Ware}}]{nardis2021stability}%
  \BibitemOpen
  \bibfield  {author} {\bibinfo {author} {\bibfnamefont {J.}~\bibnamefont {De~Nardis}}, \bibinfo {author} {\bibfnamefont {S.}~\bibnamefont {Gopalakrishnan}}, \bibinfo {author} {\bibfnamefont {R.}~\bibnamefont {Vasseur}},\ and\ \bibinfo {author} {\bibfnamefont {B.}~\bibnamefont {Ware}},\ }\bibfield  {title} {\bibinfo {title} {Stability of superdiffusion in nearly integrable spin chains},\ }\href {https://doi.org/10.1103/PhysRevLett.127.057201} {\bibfield  {journal} {\bibinfo  {journal} {Phys. Rev. Lett.}\ }\textbf {\bibinfo {volume} {127}},\ \bibinfo {pages} {057201} (\bibinfo {year} {2021})}\BibitemShut {NoStop}%
\bibitem [{\citenamefont {Schwartz}\ \emph {et~al.}(2018)\citenamefont {Schwartz}, \citenamefont {Scheuer}, \citenamefont {Tratzmiller}, \citenamefont {Müller}, \citenamefont {Chen}, \citenamefont {Dhand}, \citenamefont {Wang}, \citenamefont {Müller}, \citenamefont {Naydenov}, \citenamefont {Jelezko},\ and\ \citenamefont {Plenio}}]{schwartz_robust_2018}%
  \BibitemOpen
  \bibfield  {author} {\bibinfo {author} {\bibfnamefont {I.}~\bibnamefont {Schwartz}}, \bibinfo {author} {\bibfnamefont {J.}~\bibnamefont {Scheuer}}, \bibinfo {author} {\bibfnamefont {B.}~\bibnamefont {Tratzmiller}}, \bibinfo {author} {\bibfnamefont {S.}~\bibnamefont {Müller}}, \bibinfo {author} {\bibfnamefont {Q.}~\bibnamefont {Chen}}, \bibinfo {author} {\bibfnamefont {I.}~\bibnamefont {Dhand}}, \bibinfo {author} {\bibfnamefont {Z.-Y.}\ \bibnamefont {Wang}}, \bibinfo {author} {\bibfnamefont {C.}~\bibnamefont {Müller}}, \bibinfo {author} {\bibfnamefont {B.}~\bibnamefont {Naydenov}}, \bibinfo {author} {\bibfnamefont {F.}~\bibnamefont {Jelezko}},\ and\ \bibinfo {author} {\bibfnamefont {M.~B.}\ \bibnamefont {Plenio}},\ }\bibfield  {title} {\bibinfo {title} {Robust optical polarization of nuclear spin baths using {Hamiltonian} engineering of nitrogen-vacancy center quantum dynamics},\ }\href {https://doi.org/10.1126/sciadv.aat8978} {\bibfield  {journal} {\bibinfo  {journal} {Science Advances}\ }\textbf
  {\bibinfo {volume} {4}},\ \bibinfo {pages} {eaat8978} (\bibinfo {year} {2018})}\BibitemShut {NoStop}%
\bibitem [{\citenamefont {London}\ \emph {et~al.}(2013)\citenamefont {London}, \citenamefont {Scheuer}, \citenamefont {Cai}, \citenamefont {Schwarz}, \citenamefont {Retzker}, \citenamefont {Plenio}, \citenamefont {Katagiri}, \citenamefont {Teraji}, \citenamefont {Koizumi}, \citenamefont {Isoya}, \citenamefont {Fischer}, \citenamefont {McGuinness}, \citenamefont {Naydenov},\ and\ \citenamefont {Jelezko}}]{London:2013}%
  \BibitemOpen
  \bibfield  {author} {\bibinfo {author} {\bibfnamefont {P.}~\bibnamefont {London}}, \bibinfo {author} {\bibfnamefont {J.}~\bibnamefont {Scheuer}}, \bibinfo {author} {\bibfnamefont {J.-M.}\ \bibnamefont {Cai}}, \bibinfo {author} {\bibfnamefont {I.}~\bibnamefont {Schwarz}}, \bibinfo {author} {\bibfnamefont {A.}~\bibnamefont {Retzker}}, \bibinfo {author} {\bibfnamefont {M.~B.}\ \bibnamefont {Plenio}}, \bibinfo {author} {\bibfnamefont {M.}~\bibnamefont {Katagiri}}, \bibinfo {author} {\bibfnamefont {T.}~\bibnamefont {Teraji}}, \bibinfo {author} {\bibfnamefont {S.}~\bibnamefont {Koizumi}}, \bibinfo {author} {\bibfnamefont {J.}~\bibnamefont {Isoya}}, \bibinfo {author} {\bibfnamefont {R.}~\bibnamefont {Fischer}}, \bibinfo {author} {\bibfnamefont {L.~P.}\ \bibnamefont {McGuinness}}, \bibinfo {author} {\bibfnamefont {B.}~\bibnamefont {Naydenov}},\ and\ \bibinfo {author} {\bibfnamefont {F.}~\bibnamefont {Jelezko}},\ }\bibfield  {title} {\bibinfo {title} {Detecting and polarizing nuclear spins with double resonance on a
  single electron spin},\ }\href {https://doi.org/10.1103/PhysRevLett.111.067601} {\bibfield  {journal} {\bibinfo  {journal} {Phys. Rev. Lett.}\ }\textbf {\bibinfo {volume} {111}},\ \bibinfo {pages} {067601} (\bibinfo {year} {2013})}\BibitemShut {NoStop}%
\bibitem [{\citenamefont {Urbaszek}\ \emph {et~al.}(2013)\citenamefont {Urbaszek}, \citenamefont {Marie}, \citenamefont {Amand}, \citenamefont {Krebs}, \citenamefont {Voisin}, \citenamefont {Maletinsky}, \citenamefont {H\"ogele},\ and\ \citenamefont {Imamoglu}}]{Urbaszek:2013}%
  \BibitemOpen
  \bibfield  {author} {\bibinfo {author} {\bibfnamefont {B.}~\bibnamefont {Urbaszek}}, \bibinfo {author} {\bibfnamefont {X.}~\bibnamefont {Marie}}, \bibinfo {author} {\bibfnamefont {T.}~\bibnamefont {Amand}}, \bibinfo {author} {\bibfnamefont {O.}~\bibnamefont {Krebs}}, \bibinfo {author} {\bibfnamefont {P.}~\bibnamefont {Voisin}}, \bibinfo {author} {\bibfnamefont {P.}~\bibnamefont {Maletinsky}}, \bibinfo {author} {\bibfnamefont {A.}~\bibnamefont {H\"ogele}},\ and\ \bibinfo {author} {\bibfnamefont {A.}~\bibnamefont {Imamoglu}},\ }\bibfield  {title} {\bibinfo {title} {Nuclear spin physics in quantum dots: An optical investigation},\ }\href {https://doi.org/10.1103/RevModPhys.85.79} {\bibfield  {journal} {\bibinfo  {journal} {Rev. Mod. Phys.}\ }\textbf {\bibinfo {volume} {85}},\ \bibinfo {pages} {79} (\bibinfo {year} {2013})}\BibitemShut {NoStop}%
\bibitem [{\citenamefont {Hanson}\ \emph {et~al.}(2007)\citenamefont {Hanson}, \citenamefont {Kouwenhoven}, \citenamefont {Petta}, \citenamefont {Tarucha},\ and\ \citenamefont {Vandersypen}}]{hanson_spins_2007}%
  \BibitemOpen
  \bibfield  {author} {\bibinfo {author} {\bibfnamefont {R.}~\bibnamefont {Hanson}}, \bibinfo {author} {\bibfnamefont {L.~P.}\ \bibnamefont {Kouwenhoven}}, \bibinfo {author} {\bibfnamefont {J.~R.}\ \bibnamefont {Petta}}, \bibinfo {author} {\bibfnamefont {S.}~\bibnamefont {Tarucha}},\ and\ \bibinfo {author} {\bibfnamefont {L.~M.~K.}\ \bibnamefont {Vandersypen}},\ }\bibfield  {title} {\bibinfo {title} {Spins in few-electron quantum dots},\ }\href {https://doi.org/10.1103/RevModPhys.79.1217} {\bibfield  {journal} {\bibinfo  {journal} {Reviews of Modern Physics}\ }\textbf {\bibinfo {volume} {79}},\ \bibinfo {pages} {1217} (\bibinfo {year} {2007})}\BibitemShut {NoStop}%
\bibitem [{\citenamefont {Dukelsky}\ \emph {et~al.}(2004)\citenamefont {Dukelsky}, \citenamefont {Pittel},\ and\ \citenamefont {Sierra}}]{dukelsky_colloquim_2004}%
  \BibitemOpen
  \bibfield  {author} {\bibinfo {author} {\bibfnamefont {J.}~\bibnamefont {Dukelsky}}, \bibinfo {author} {\bibfnamefont {S.}~\bibnamefont {Pittel}},\ and\ \bibinfo {author} {\bibfnamefont {G.}~\bibnamefont {Sierra}},\ }\bibfield  {title} {\bibinfo {title} {Colloquium: Exactly solvable richardson-gaudin models for many-body quantum systems},\ }\href {https://doi.org/10.1103/RevModPhys.76.643} {\bibfield  {journal} {\bibinfo  {journal} {Rev. Mod. Phys.}\ }\textbf {\bibinfo {volume} {76}},\ \bibinfo {pages} {643} (\bibinfo {year} {2004})}\BibitemShut {NoStop}%
\bibitem [{\citenamefont {Yuzbashyan}\ \emph {et~al.}(2005)\citenamefont {Yuzbashyan}, \citenamefont {Altshuler}, \citenamefont {Kuznetsov},\ and\ \citenamefont {Enolskii}}]{yuzbashyan_solution_2005}%
  \BibitemOpen
  \bibfield  {author} {\bibinfo {author} {\bibfnamefont {E.~A.}\ \bibnamefont {Yuzbashyan}}, \bibinfo {author} {\bibfnamefont {B.~L.}\ \bibnamefont {Altshuler}}, \bibinfo {author} {\bibfnamefont {V.~B.}\ \bibnamefont {Kuznetsov}},\ and\ \bibinfo {author} {\bibfnamefont {V.~Z.}\ \bibnamefont {Enolskii}},\ }\bibfield  {title} {\bibinfo {title} {Solution for the dynamics of the bcs and central spin problems},\ }\href {https://doi.org/10.1088/0305-4470/38/36/003} {\bibfield  {journal} {\bibinfo  {journal} {Journal of Physics A: Mathematical and General}\ }\textbf {\bibinfo {volume} {38}},\ \bibinfo {pages} {7831} (\bibinfo {year} {2005})}\BibitemShut {NoStop}%
\bibitem [{\citenamefont {Dobrzyniecki}\ and\ \citenamefont {Tomza}(2023)}]{dobrzyniecki_quantum_2023}%
  \BibitemOpen
  \bibfield  {author} {\bibinfo {author} {\bibfnamefont {J.}~\bibnamefont {Dobrzyniecki}}\ and\ \bibinfo {author} {\bibfnamefont {M.}~\bibnamefont {Tomza}},\ }\bibfield  {title} {\bibinfo {title} {Quantum simulation of the central spin model with a rydberg atom and polar molecules in optical tweezers},\ }\href {https://doi.org/10.1103/PhysRevA.108.052618} {\bibfield  {journal} {\bibinfo  {journal} {Phys. Rev. A}\ }\textbf {\bibinfo {volume} {108}},\ \bibinfo {pages} {052618} (\bibinfo {year} {2023})}\BibitemShut {NoStop}%
\bibitem [{\citenamefont {Kittel}\ and\ \citenamefont {Shore}(1965)}]{kittel_development_1965}%
  \BibitemOpen
  \bibfield  {author} {\bibinfo {author} {\bibfnamefont {C.}~\bibnamefont {Kittel}}\ and\ \bibinfo {author} {\bibfnamefont {H.}~\bibnamefont {Shore}},\ }\bibfield  {title} {\bibinfo {title} {Development of a phase transition for a rigorously solvable many-body system},\ }\href {https://doi.org/10.1103/PhysRev.138.A1165} {\bibfield  {journal} {\bibinfo  {journal} {Phys. Rev.}\ }\textbf {\bibinfo {volume} {138}},\ \bibinfo {pages} {A1165} (\bibinfo {year} {1965})}\BibitemShut {NoStop}%
\bibitem [{\citenamefont {Magyari}\ \emph {et~al.}(1987)\citenamefont {Magyari}, \citenamefont {Thomas}, \citenamefont {Weber}, \citenamefont {Kaufman},\ and\ \citenamefont {M\"uller}}]{magyari_integrable_1987}%
  \BibitemOpen
  \bibfield  {author} {\bibinfo {author} {\bibfnamefont {E.}~\bibnamefont {Magyari}}, \bibinfo {author} {\bibfnamefont {H.}~\bibnamefont {Thomas}}, \bibinfo {author} {\bibfnamefont {R.}~\bibnamefont {Weber}}, \bibinfo {author} {\bibfnamefont {C.}~\bibnamefont {Kaufman}},\ and\ \bibinfo {author} {\bibfnamefont {G.}~\bibnamefont {M\"uller}},\ }\bibfield  {title} {\bibinfo {title} {Integrable and nonintegrable classical spin clusters},\ }\href {https://doi.org/10.1007/BF01303725} {\bibfield  {journal} {\bibinfo  {journal} {Zeitschrift f\"ur Physik B Condensed Matter}\ }\textbf {\bibinfo {volume} {65}},\ \bibinfo {pages} {363} (\bibinfo {year} {1987})}\BibitemShut {NoStop}%
\bibitem [{\citenamefont {Liu}\ and\ \citenamefont {M\"uller}(1990)}]{liu_infinte_1990}%
  \BibitemOpen
  \bibfield  {author} {\bibinfo {author} {\bibfnamefont {J.-M.}\ \bibnamefont {Liu}}\ and\ \bibinfo {author} {\bibfnamefont {G.}~\bibnamefont {M\"uller}},\ }\bibfield  {title} {\bibinfo {title} {Infinite-temperature dynamics of the equivalent-neighbor xyz model},\ }\href {https://doi.org/10.1103/PhysRevA.42.5854} {\bibfield  {journal} {\bibinfo  {journal} {Phys. Rev. A}\ }\textbf {\bibinfo {volume} {42}},\ \bibinfo {pages} {5854} (\bibinfo {year} {1990})}\BibitemShut {NoStop}%
\bibitem [{\citenamefont {Wu}\ \emph {et~al.}(2020)\citenamefont {Wu}, \citenamefont {Guan},\ and\ \citenamefont {Links}}]{wu_separable_2020}%
  \BibitemOpen
  \bibfield  {author} {\bibinfo {author} {\bibfnamefont {N.}~\bibnamefont {Wu}}, \bibinfo {author} {\bibfnamefont {X.-W.}\ \bibnamefont {Guan}},\ and\ \bibinfo {author} {\bibfnamefont {J.}~\bibnamefont {Links}},\ }\bibfield  {title} {\bibinfo {title} {Separable and entangled states in the high-spin $xx$ central spin model},\ }\href {https://doi.org/10.1103/PhysRevB.101.155145} {\bibfield  {journal} {\bibinfo  {journal} {Phys. Rev. B}\ }\textbf {\bibinfo {volume} {101}},\ \bibinfo {pages} {155145} (\bibinfo {year} {2020})}\BibitemShut {NoStop}%
\bibitem [{\citenamefont {De~Nadai}\ \emph {et~al.}(2024)\citenamefont {De~Nadai}, \citenamefont {Maestracci},\ and\ \citenamefont {Faribault}}]{nadai_integrability_2024}%
  \BibitemOpen
  \bibfield  {author} {\bibinfo {author} {\bibfnamefont {E.}~\bibnamefont {De~Nadai}}, \bibinfo {author} {\bibfnamefont {N.}~\bibnamefont {Maestracci}},\ and\ \bibinfo {author} {\bibfnamefont {A.}~\bibnamefont {Faribault}},\ }\bibfield  {title} {\bibinfo {title} {Integrability and dark states of the xx spin-1 central spin model in a transverse field},\ }\href {https://doi.org/10.1103/PhysRevB.110.205427} {\bibfield  {journal} {\bibinfo  {journal} {Phys. Rev. B}\ }\textbf {\bibinfo {volume} {110}},\ \bibinfo {pages} {205427} (\bibinfo {year} {2024})}\BibitemShut {NoStop}%
\bibitem [{\citenamefont {Tonder}\ and\ \citenamefont {Links}(2025)}]{tonder_dark_2025}%
  \BibitemOpen
  \bibfield  {author} {\bibinfo {author} {\bibfnamefont {J.~v.}\ \bibnamefont {Tonder}}\ and\ \bibinfo {author} {\bibfnamefont {J.}~\bibnamefont {Links}},\ }\bibfield  {title} {\bibinfo {title} {Dark states in an integrable xy central spin model},\ }\href {https://doi.org/10.1088/1742-5468/adde3d} {\bibfield  {journal} {\bibinfo  {journal} {Journal of Statistical Mechanics: Theory and Experiment}\ }\textbf {\bibinfo {volume} {2025}},\ \bibinfo {pages} {063104} (\bibinfo {year} {2025})}\BibitemShut {NoStop}%
\bibitem [{\citenamefont {Kubo}(1957)}]{kubo_statistical_1957}%
  \BibitemOpen
  \bibfield  {author} {\bibinfo {author} {\bibfnamefont {R.}~\bibnamefont {Kubo}},\ }\bibfield  {title} {\bibinfo {title} {Statistical-mechanical theory of irreversible processes. i. general theory and simple applications to magnetic and conduction problems},\ }\href {https://doi.org/10.1143/JPSJ.12.570} {\bibfield  {journal} {\bibinfo  {journal} {Journal of the Physical Society of Japan}\ }\textbf {\bibinfo {volume} {12}},\ \bibinfo {pages} {570} (\bibinfo {year} {1957})}\BibitemShut {NoStop}%
\bibitem [{\citenamefont {Arnold}(1989)}]{Arnold1989_classicalmechanics}%
  \BibitemOpen
  \bibfield  {author} {\bibinfo {author} {\bibfnamefont {V.~I.}\ \bibnamefont {Arnold}},\ }\href {https://doi.org/10.1007/978-1-4757-2063-1} {\emph {\bibinfo {title} {{Mathematical Methods of Classical Mechanics}}}},\ \bibinfo {edition} {2nd}\ ed.,\ {Graduate Texts in Mathematics}\ (\bibinfo  {publisher} {{Springer New York}},\ \bibinfo {year} {1989})\BibitemShut {NoStop}%
\bibitem [{\citenamefont {Fine}\ \emph {et~al.}(2014)\citenamefont {Fine}, \citenamefont {Elsayed}, \citenamefont {Kropf},\ and\ \citenamefont {de~Wijn}}]{Fine_2014}%
  \BibitemOpen
  \bibfield  {author} {\bibinfo {author} {\bibfnamefont {B.~V.}\ \bibnamefont {Fine}}, \bibinfo {author} {\bibfnamefont {T.~A.}\ \bibnamefont {Elsayed}}, \bibinfo {author} {\bibfnamefont {C.~M.}\ \bibnamefont {Kropf}},\ and\ \bibinfo {author} {\bibfnamefont {A.~S.}\ \bibnamefont {de~Wijn}},\ }\bibfield  {title} {\bibinfo {title} {Absence of exponential sensitivity to small perturbations in nonintegrable systems of spins 1/2},\ }\href {https://doi.org/10.1103/PhysRevE.89.012923} {\bibfield  {journal} {\bibinfo  {journal} {Phys. Rev. E}\ }\textbf {\bibinfo {volume} {89}},\ \bibinfo {pages} {012923} (\bibinfo {year} {2014})}\BibitemShut {NoStop}%
\bibitem [{\citenamefont {Kukuljan}\ \emph {et~al.}(2017)\citenamefont {Kukuljan}, \citenamefont {Grozdanov},\ and\ \citenamefont {Prosen}}]{Kukuljan_2017}%
  \BibitemOpen
  \bibfield  {author} {\bibinfo {author} {\bibfnamefont {I.}~\bibnamefont {Kukuljan}}, \bibinfo {author} {\bibfnamefont {S.}~\bibnamefont {Grozdanov}},\ and\ \bibinfo {author} {\bibfnamefont {T.}~\bibnamefont {Prosen}},\ }\bibfield  {title} {\bibinfo {title} {Weak quantum chaos},\ }\href {https://doi.org/10.1103/PhysRevB.96.060301} {\bibfield  {journal} {\bibinfo  {journal} {Phys. Rev. B}\ }\textbf {\bibinfo {volume} {96}},\ \bibinfo {pages} {060301} (\bibinfo {year} {2017})}\BibitemShut {NoStop}%
\bibitem [{\citenamefont {Davidson}\ \emph {et~al.}(2017)\citenamefont {Davidson}, \citenamefont {Sels},\ and\ \citenamefont {Polkovnikov}}]{Davidson_2017}%
  \BibitemOpen
  \bibfield  {author} {\bibinfo {author} {\bibfnamefont {S.}~\bibnamefont {Davidson}}, \bibinfo {author} {\bibfnamefont {D.}~\bibnamefont {Sels}},\ and\ \bibinfo {author} {\bibfnamefont {A.}~\bibnamefont {Polkovnikov}},\ }\bibfield  {title} {\bibinfo {title} {Semiclassical approach to dynamics of interacting fermions},\ }\href {https://doi.org/10.1016/j.aop.2017.07.003} {\bibfield  {journal} {\bibinfo  {journal} {Annals of Physics}\ }\textbf {\bibinfo {volume} {384}},\ \bibinfo {pages} {128–141} (\bibinfo {year} {2017})}\BibitemShut {NoStop}%
\bibitem [{\citenamefont {Schmitt}\ \emph {et~al.}(2019)\citenamefont {Schmitt}, \citenamefont {Sels}, \citenamefont {Kehrein},\ and\ \citenamefont {Polkovnikov}}]{Schmitt_2019}%
  \BibitemOpen
  \bibfield  {author} {\bibinfo {author} {\bibfnamefont {M.}~\bibnamefont {Schmitt}}, \bibinfo {author} {\bibfnamefont {D.}~\bibnamefont {Sels}}, \bibinfo {author} {\bibfnamefont {S.}~\bibnamefont {Kehrein}},\ and\ \bibinfo {author} {\bibfnamefont {A.}~\bibnamefont {Polkovnikov}},\ }\bibfield  {title} {\bibinfo {title} {Semiclassical echo dynamics in the sachdev-ye-kitaev model},\ }\href {https://doi.org/10.1103/PhysRevB.99.134301} {\bibfield  {journal} {\bibinfo  {journal} {Phys. Rev. B}\ }\textbf {\bibinfo {volume} {99}},\ \bibinfo {pages} {134301} (\bibinfo {year} {2019})}\BibitemShut {NoStop}%
\bibitem [{\citenamefont {de~Wijn}\ \emph {et~al.}(2013)\citenamefont {de~Wijn}, \citenamefont {Hess},\ and\ \citenamefont {Fine}}]{wijn_lyapunov_2013}%
  \BibitemOpen
  \bibfield  {author} {\bibinfo {author} {\bibfnamefont {A.~S.}\ \bibnamefont {de~Wijn}}, \bibinfo {author} {\bibfnamefont {B.}~\bibnamefont {Hess}},\ and\ \bibinfo {author} {\bibfnamefont {B.~V.}\ \bibnamefont {Fine}},\ }\bibfield  {title} {\bibinfo {title} {Lyapunov instabilities in lattices of interacting classical spins at infinite temperature},\ }\href {https://doi.org/10.1088/1751-8113/46/25/254012} {\bibfield  {journal} {\bibinfo  {journal} {Journal of Physics A: Mathematical and Theoretical}\ }\textbf {\bibinfo {volume} {46}},\ \bibinfo {pages} {254012} (\bibinfo {year} {2013})}\BibitemShut {NoStop}%
\bibitem [{\citenamefont {Meiss}\ and\ \citenamefont {Ott}(1985)}]{meiss_markov-tree_1985}%
  \BibitemOpen
  \bibfield  {author} {\bibinfo {author} {\bibfnamefont {J.~D.}\ \bibnamefont {Meiss}}\ and\ \bibinfo {author} {\bibfnamefont {E.}~\bibnamefont {Ott}},\ }\bibfield  {title} {\bibinfo {title} {Markov-{Tree} {Model} of {Intrinsic} {Transport} in {Hamiltonian} {Systems}},\ }\href {https://doi.org/10.1103/PhysRevLett.55.2741} {\bibfield  {journal} {\bibinfo  {journal} {Physical Review Letters}\ }\textbf {\bibinfo {volume} {55}},\ \bibinfo {pages} {2741} (\bibinfo {year} {1985})}\BibitemShut {NoStop}%
\bibitem [{\citenamefont {Srivastava}\ \emph {et~al.}(1988)\citenamefont {Srivastava}, \citenamefont {Kaufman}, \citenamefont {Müller}, \citenamefont {Weber},\ and\ \citenamefont {Thomas}}]{srivastava_integrable_1988}%
  \BibitemOpen
  \bibfield  {author} {\bibinfo {author} {\bibfnamefont {N.}~\bibnamefont {Srivastava}}, \bibinfo {author} {\bibfnamefont {C.}~\bibnamefont {Kaufman}}, \bibinfo {author} {\bibfnamefont {G.}~\bibnamefont {Müller}}, \bibinfo {author} {\bibfnamefont {R.}~\bibnamefont {Weber}},\ and\ \bibinfo {author} {\bibfnamefont {H.}~\bibnamefont {Thomas}},\ }\bibfield  {title} {\bibinfo {title} {Integrable and {Nonintegrable} {Classical} {Spin} {Clusters}: {Trajectories} and {Geometric} {Structure} of {Invariants} and {Geometric} {Structure} of {Invariants}},\ }\href {https://doi.org/10.1007/BF01318307} {\bibfield  {journal} {\bibinfo  {journal} {Zeitschrift für Physik B Condensed Matter}\ }\textbf {\bibinfo {volume} {70}},\ \bibinfo {pages} {251} (\bibinfo {year} {1988})}\BibitemShut {NoStop}%
\bibitem [{\citenamefont {Johnson}(2014)}]{Soboljl}%
  \BibitemOpen
  \bibfield  {author} {\bibinfo {author} {\bibfnamefont {S.~G.}\ \bibnamefont {Johnson}},\ }\href {https://github.com/JuliaMath/Sobol.jl} {\bibinfo {title} {Sobol.jl}} (\bibinfo {year} {2014})\BibitemShut {NoStop}%
\bibitem [{Cha(2025)}]{ChaosTools}%
  \BibitemOpen
  \href@noop {} {\bibinfo {title} {{ChaosTools.jl}}},\ \bibinfo {howpublished} {\url{https://github.com/JuliaDynamics/ChaosTools.jl}} (\bibinfo {year} {2025})\BibitemShut {NoStop}%
\bibitem [{\citenamefont {Datseris}(2018)}]{Datseris2018}%
  \BibitemOpen
  \bibfield  {author} {\bibinfo {author} {\bibfnamefont {G.}~\bibnamefont {Datseris}},\ }\bibfield  {title} {\bibinfo {title} {Dynamicalsystems.jl: A julia software library for chaos and nonlinear dynamics},\ }\href {https://doi.org/10.21105/joss.00598} {\bibfield  {journal} {\bibinfo  {journal} {Journal of Open Source Software}\ }\textbf {\bibinfo {volume} {3}},\ \bibinfo {pages} {598} (\bibinfo {year} {2018})}\BibitemShut {NoStop}%
\bibitem [{\citenamefont {Datseris}\ and\ \citenamefont {Parlitz}(2022)}]{DatserisParlitz2022}%
  \BibitemOpen
  \bibfield  {author} {\bibinfo {author} {\bibfnamefont {G.}~\bibnamefont {Datseris}}\ and\ \bibinfo {author} {\bibfnamefont {U.}~\bibnamefont {Parlitz}},\ }\href {https://doi.org/10.1007/978-3-030-91032-7} {\emph {\bibinfo {title} {Nonlinear dynamics: A concise introduction interlaced with code}}}\ (\bibinfo  {publisher} {Springer Nature},\ \bibinfo {address} {Cham, Switzerland},\ \bibinfo {year} {2022})\BibitemShut {NoStop}%
\end{thebibliography}%

\appendix

\section{Superintegrability in the XX central spin model at the zero energy}\label{app:regular_CS}

We prove that motion in the central spin model with XX interactions defined in \autoref{eq:CS} in the main text is regular in the zero-energy manifold ($E=0$). Specifically, we show that this manifold is superintegrable by identifying $2L-1$ independent integrals of motion.

To establish integrability in a Hamiltonian system of $L$ spins at least $L$ (functionally) independent integrals of motion $I_j$:
\begin{equation}
    \dv{I_j}{t} = \poissonbracket{I_j}{\HCS} = 0
    \label{Eq:DefnInteg}
\end{equation}
that are mutually in involution $\poissonbracket{I_j}{I_i}=0$ are required. The Liouville-Arnold theorem then guarantees that every trajectory is regular~\cite{Arnold1989_classicalmechanics}.

A model is considered superintegrable if additional independent integrals of motion are present~\cite{Mishchenko1978superintegrable,Bolsinov2003noncommutative,miller_classical_2013}. Each such additional integral reduces the phase space dimension available to the system’s dynamics, so that the trajectories in a system with $L+M$ independent integrals are restricted to a $L-M$-dimensional manifold. The $E=0$ manifold of the central spin model is an extreme case of superintegrability, also called \textit{maximally superintegrable}, with $2L-1$ integrals of motion restricting trajectories to one-dimensional closed orbits.

We start by constructing one conserved quantity $I_j$ per spin. Consequently, the motion of each spin is parameterized by a single angle, $\alpha_j(t)$, rather than the two angles needed to specify a point on the unit sphere, and each spin's trajectory is confined to a circle on the sphere. 
We note that these angles do not form action-angle variables with the conserved quantities \(I_j\) because they do not evolve linearly in time.
We then construct $L-2$ additional, independent conserved quantities $Q_{j}$ for $j\geq 2$ from the obtained parameterization by deriving the equations of motion for $\alpha_j$. 
Together with the total magnetization $\Sztot$, the set $\left\{I_0,\dots,I_{L-1},Q_2,\dots,Q_{L-1},\Sztot\right\}$ contains the $2L-1$ independent integrals of motion, when restricted to the $E = 0$ manifold.

The \emph{mean field} experienced by the central spin, $\BperpVec =\begin{pmatrix}B^x, & B^y, & 0\end{pmatrix}= \sum_{j>0}g_j\begin{pmatrix}S_j^x, & S_j^y, & 0\end{pmatrix}$, plays a crucial role in proving the superintegrability of the model. In terms of the mean field, the Hamiltonian is:
\begin{equation}
    \HCS = \SperpVec \vdot \BperpVec.\label{eq:two_spins}
\end{equation}
It follows that $E=0$ enforces that the transverse mean field generated by the environment spins is orthogonal to the transverse components of the central spin at all times.

We find that the following quantity is conserved within the zero-energy shell 
\begin{align}
\label{eq:I0iom_rep}
I_0=\frac{S_0^x}{S_0^y}=\cot(\varphi_0).
\end{align} 
Using the equations of motion, $\dot{S_0^x}=S_0^zB^y$ and $\dot{S_0^y}=-S_0^z B^x$, we find,
\begin{align}
     \dot{I_0} = \frac{\dot{S_0^x}S_0^y-S_0^x\dot{S_0^y}}{{S_0^y}^2}= \frac{S_0^z}{{S_0^y}^2} \left(S_0^yB^y+S_0^xB^x\right)= \frac{S_0^z}{{S_0^y}^2} \HCS.
\end{align}
Thus,
\begin{equation}
    \dot{I_0} = 0 \text{ for } E=0,  \label{eq:I0}
\end{equation}
as required for a conserved quantity. The conservation of $I_0$ is equivalent to the conservation of $\varphi_0$ mod $\pi$.

Accounting for the constant magnitude $\vec{S}_0 \vdot \vec{S}_0=1$, the motion of the central spin is confined to a great circle on the sphere perpendicular to the axis of the mean field. The motion is parameterized by $\alpha_0(t)$, the angle between $\vec{S}_0$ and the $z$-axis,
\begin{equation}
    \vec{S}_0(t) = \frac{\sin(\alpha_0(t))}{\sqrt{1+I_0^2}}\begin{pmatrix} I_0 \\ 1 \\ 0\end{pmatrix} + \cos(\alpha_0(t))\begin{pmatrix} 0 \\ 0 \\ 1\end{pmatrix}.
    \label{eq:alpha0}
\end{equation}

The angle $\alpha_0(t)$ does not increase uniformly in time; indeed, there are trajectories in which $\vb{S}_0$ only explores a part of the great circle for all times. 

The motion of every environment spin is also restricted to a circle on the unit sphere.
Each environment spin experiences a field in the $xy$-plane along the constant axis determined by $(S_0^x, S_0^y, 0) \propto (I_0, 1, 0)$. Although the magnitude and sign of the field vary with time, the direction is unchanged. Consequently, the angle between $\vec{S}_j$ and the constant axis is a constant of the motion:
\begin{equation}
\label{Eq:Ijiom}
    I_j = \begin{pmatrix} I_0 \\ 1 \\ 0\end{pmatrix}\vdot\begin{pmatrix} S^x_j \\ S^y_j \\ S^z_j \end{pmatrix} = {I_0S_j^x+ S_j^y} .
\end{equation}
We can explicitly check that $I_j$ is conserved using $\dot{S_j^x}=g_j S^y_0S_j^z$ and $\dot{S_j^y}=-g_j S^x_0S_j^z$:
\begin{align}
     \dot{{I}_j} = \frac{S_0^zS_j^x}{{S_0^y}^2}\HCS = 0 \text{ for } E = 0 . \label{eq:Ij}
\end{align}

As in the case of the central spin, each environment spin's motion is parameterized by a single angle, $\alpha_j(t)$. This is the angle between the $z$-axis and the spin's projection on the plane perpendicular to $(I_0,1,0)$:  
\begin{align}
    \vec{S}_j(t) =& \frac{I_j}{{1+I_0^2}}\begin{pmatrix}  I_0\\ 1\\ 0 \end{pmatrix}+ \cos(\alpha_j(t))\sqrt{\frac{1+I_0^2-I_j^2}{1+I_0^2}}\begin{pmatrix} 0 \\ 0\\ 1 \end{pmatrix}\nonumber\\
    +& \sin(\alpha_j(t))\frac{\sqrt{1+I_0^2-I_j^2}}{1+I_0^2}\begin{pmatrix} 1 \\ -I_0\\ 0 \end{pmatrix}.\label{eq:Sj}
\end{align}

Using $\dot{S_j^x}=g_j S^y_0S_j^z$ and the derived parameterization in \autoref{eq:Sj}, we obtain the equation of motion for $\alpha_j(t)$:
\begin{align}
    \dot{S}_j^x = g_j S_0^yS_j^z = \frac{\sqrt{1+I_0^2-I_j^2}}{1+I_0^2}\cos(\alpha_j)\dot{\alpha_j}.
\end{align}
By further substituting $S_j^z$ from \autoref{eq:Sj}, we arrive at
\begin{align}
    \dot{\alpha}_j = g_j \sqrt{1+I_0^2}S_0^y= g_j \sqrt{(S_0^x)^2+(S_0^y)^2}=g_j\sin(\alpha_0).\label{eq:dot_alpha_j}
\end{align}
Thus, the dynamics of every environment spin's angle is determined by the angle of the central spin. 
\autoref{eq:dot_alpha_j} immediately identifies the remaining $L-2$ integrals of motion:
\begin{align}
    Q_{j} &= g_1\alpha_j-g_j\alpha_1\text{ for } j\geq 2\label{eq:Q_j}\\
    \dot{Q}_{j} &= g_1g_j\sin(\alpha_0)- g_1g_j\sin(\alpha_0)=0.
\end{align}
We have defined $Q_j$ with respect to the first environment spin, but $g_k \alpha_j - g_j \alpha_k$ is conserved for any two environment spins $j,k$.

For completeness, we write the equation of motion for $\alpha_0$, which defines the motion of the central spin and, therefore, governs the entire system. Using $\dot{S}_0^x=\sum_jg_jS_0^zS_j^y$ and the parameterization in \autoref{eq:alpha0}:
\begin{align}
    \dot{S}_j^x &= \sum_jg_jS_0^zS_j^y = \frac{I_0}{\sqrt{1+I_0^2}}\cos(\alpha_0)\dot{\alpha_0}.\\
    \dot{\alpha}_0 &= \frac{\sqrt{1+I_0^2}}{I_0}\sum_jg_jS_j^y\nonumber\\
    &= \frac{\sqrt{1+I_0^2}}{I_0}\sum_jg_j\left(\frac{I_j}{{1+I_0^2}}-\frac{I_0\sqrt{1+I_0^2-I_j^2}}{1+I_0^2}\sin(\alpha_j)\right)\nonumber\\
    &= -\sum_jg_j\sqrt{\frac{{1+I_0^2-I_j^2}}{1+I_0^2}}\sin(\alpha_j)\nonumber\\
    &= -\sum_jg_j\sqrt{\frac{{1+I_0^2-I_j^2}}{1+I_0^2}}\sin\left(\frac{Q_j+g_j\alpha_1}{g_1}.\right)
    \label{eq:dot_alpha_0}
\end{align}
Above, we have used \autoref{eq:Q_j} and $\sum_jg_jI_j=0$, which holds almost everywhere with $E=0$ since 
\begin{equation}
    \HCS = \frac{\sin(\alpha_0)}{\sqrt{1+I_0^2}} \sum_{j=1}^{L-1} g_j I_j.
\end{equation}
We emphasize that the conservation of $I_j, Q_j$ requires $E=0$.

To demonstrate the functional independence of the conserved quantities, we construct a diffeomorphism between a dense open subset of the old coordinates $\vb{S}$ and a dense open subset of the new coordinates $\vb{T}:=\left(\alpha_1,\varphi_0,I_1,\dots,I_{L-1},Q_2,\dots,Q_{L-1},\Sztot\right)$, where $\varphi_0$ is the azimuthal angle of the central spin. Given a spin configuration $\vb{S}$, we obtain all $I_j$ and $\alpha_j$ (and $\Sztot$), which in turn define $Q_j$, and thus a point in $\vb{T}$. The inverse map, from $\vb{T}$ to $\vb{S}$, is obtained in the following way. From $\varphi_0$, we obtain $I_0=\cos(\varphi_0)/\sin(\varphi_0)$. \autoref{eq:Q_j} obtains $\alpha_j$ for every environment spin with $j\geq 2$, given $\alpha_1$ and $Q_j$. \autoref{eq:Sj} then determines $\vb{S}_j$ for $j\geq 1$. Lastly, we can use the total magnetization to obtain the polar angle of the central spin, $\theta_0= \arccos\left( \Sztot-\sum_{j\geq 1}\cos(\theta_j)\right)$, and thus $\vb{S}_0$.

In the $\vb{T}$ coordinates, only $\alpha_1$ and $\varphi_0$ vary in time. However, $\varphi_0$ only takes two discrete values within $[0,2\pi)$: $\varphi_0 = \cot^{-1}(I_0) \pm n\pi$ for $n\in\mathbb{Z}$, restricting the dynamics of $\alpha_1\in [0,2\pi)$ and $\varphi_0\in\left\{\cot^{-1}(I_0), \cot^{-1}(I_0) +\pi\right\}$. The two discrete values taken by $\varphi$ distinguishes between two points in phase space where $\dot{\alpha}_1>0$ and $\dot{\alpha}_1<0$. This can be seen from \autoref{eq:dot_alpha_j} since $\varphi_0$ changes when $\alpha_0$ changes sign.  
Thus, trajectories are restricted to one-dimensional manifolds, which implies that they are periodic for almost all initial conditions. 
The variable $\alpha$, as used in the main text, refers to a parameterization of the orbit in terms of $\alpha_1$ and the set $\left\{\cot^{-1}(I_0), \cot^{-1}(I_0) +\pi\right\}$.

\section{Supplementary data for the perturbed XX central spin model at zero energy}\label{app:supp_CS}

This section provides supplementary data supporting the claims made in \autoref{sec:centralspin} of the main text. Specifically, we demonstrate:
$(i)$ that perturbed trajectories remain close to the integrable dynamics for the majority of phase space;
$(ii)$ additional evidence that the decay of the autocorrelator is driven by large changes associated to the thin chaotic manifold defined by \(\eta \lesssim \vert\delta\vert\); and additionally
$(iii)$ that the time required to recover the microcanonical distribution of an integral of motion exceeds the thermalization time $\Tth$ extracted from the decay of the autocorrelator. 

\begin{figure}[t]
    \centering
    \includegraphics[width=\columnwidth]{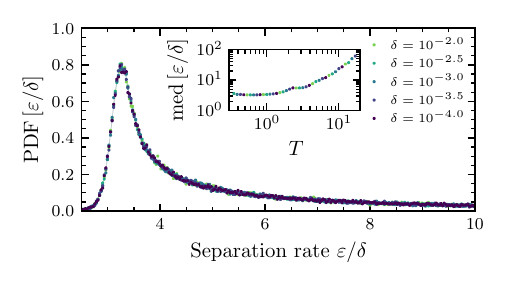}
    \caption[]{Probability distribution of the separation rate after one period, defined as $\varepsilon := \|\vb{S}_\delta(0) - \vb{S}_\delta(T)\|_2 / T$, rescaled by the perturbation strength $\delta$. The trajectory $\vb{S}_\delta$ is evolved for a time $T$, which corresponds to the period of the unperturbed ($\delta = 0$) dynamics. Shadings for each $\delta$ in the main panel represent one standard deviation error bars and are barely visible. The inset shows the rescaled separation rate as a function of $T$. Here, $\mathrm{med}$ denotes the median of all samples within a given time window. We have used at least $10^5$ samples and set $L=46$.} 
    \label{fig:error_after_T_CS}
\end{figure}

\paragraph*{(i)}
Perturbed trajectories remain close to the integrable dynamics, and unperturbed trajectories with longer periods are less stable than trajectories with shorter periods. We sample points within the zero-energy shell of \( \HCS\) and determine the period $T$. We then evolve initial conditions sampled from this trajectory for time \(T\) under the perturbed Hamiltonian \(\HCS + \delta V\). Thereby, we define the distance of the initial and final points divided by \(T\) to be the average separation rate: \( \varepsilon := \|\vb{S}_\delta(0) - \vb{S}_\delta(T)\|_2/T \).
\autoref{fig:error_after_T_CS} shows the probability distribution of the separation rate divided by the perturbation strength, $\mathrm{PDF}[\varepsilon / \delta]$. The data collapses across several orders of magnitude in $\delta$, indicating that the rate scales linearly with the perturbation strength. The inset displays the dependence of the separation rate on the period $T$, which also collapses under the same rescaling. We observe that the separation rate increases with $T$.

\paragraph*{(ii)}
We associate the encounters with the chaotic manifold as the underlying mechanism behind the broad distribution of decay times.
To illustrate this, we evaluate numerically the change in the autocorrelator, $\Delta C_{\tau_0}(T)$, over a fixed time interval \( T \) for various values of \( \delta \) and different trajectories:
\begin{align}
     \Delta C_{\tau_0}(T) := {\frac{{\Delta\tau_0(T)\tau_0(0)}}{\expval{\tau_0(0)\tau_0(0)}}}.\label{eq:Ctau_0_short_time}
\end{align}
Here, $\Delta\tau_0(T)=\tau_0(T)-\tau_0(0)$ refers to the change in $\tau_0$ over $T$ and the autocorrelator is $C_{\tau_0}(T)=1+\Delta C_{\tau_0}(T)$. $\tau_0$ is bounded within $[-1,1]$, symmetric around zero, and $\expval{\tau_0(0)\tau_0(0)}$ evaluates to $1/2$. Crucially, $\Delta C_{\tau_0}$ can take values $[-4,1/2]$.

\begin{figure}[t]
    \centering
    \includegraphics[width=\columnwidth]{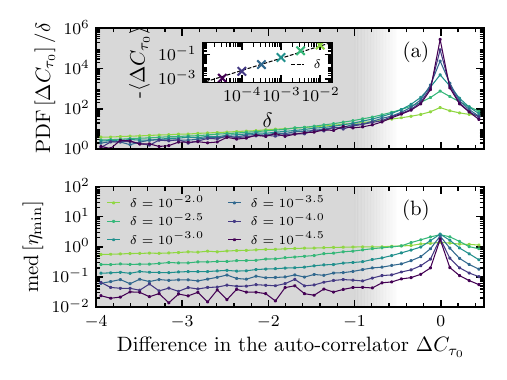}
	\caption{Numerical evidence that the decay of the autocorrelator in the central spin model originates from rare encounters with chaotic manifolds. The $x$ axis shows the change in the autocorrelator \( \Delta C_{\tau_0} \) of \( \tau_0 \) as written in \autoref{eq:Ctau_0_short_time}. The simulation time is \( T = 100/J \). Panel (a) shows the probability density of \( \Delta C_{\tau_0} \), rescaled by \( \delta^{-1} \). The inset displays the mean of the unscaled distribution as a function of \( \delta \), revealing a linear dependence.  Panel (b) shows the median of the minimal value \( \etamin \) observed within the simulation as a function of \( \Delta C_{\tau_0} \). We used $10^6$ samples, and the system contained $L=46$ spins.}\label{fig:decay_CS}
\end{figure}

The probability density of $\Delta C_{\tau_0}$ rescaled by $\vert\delta\vert^{-1}$ is shown in \figref{fig:decay_CS}{a}. The inset shows the mean of the distribution, revealing a linear decay with $\vert\delta\vert$. While the bulk of the distribution is symmetrically centered around the peak, we find a long tail of the distribution, whose mass scales as $\vert\delta\vert$, as indicated by the data collapse. The height scaling, along with the asymmetric support of the distribution of $[-4,1/2]$, results in a linear dependence of the mean on $\vert\delta\vert$.

The tail arises from rare,  $\mathcal{O}(1)$ deviations in $\tau_0$, which occur only within the chaotic manifold. For all trajectories with a given $\Delta C_{\tau_0}$, we compute $\eta_\text{min}$ the minimal value of $\eta$ (defined in \autoref{eq:eta} in the main text) within the evolution until time $T$. Then we plot the median of $\etamin$ as a function of $\Delta C_{\tau_0}$ in \figref{fig:decay_CS}{b}. We find that trajectories near the peak of the distribution (with small $|\Delta C_{\tau_0}|$) tend to exhibit large values of $\eta$, consistent with regular behavior. In contrast, the tails of the distribution correspond to trajectories with small values of $\eta$, indicating that they have traversed the chaotic manifold.

\begin{figure*}[t]
    \centering
    \includegraphics[width=\textwidth]{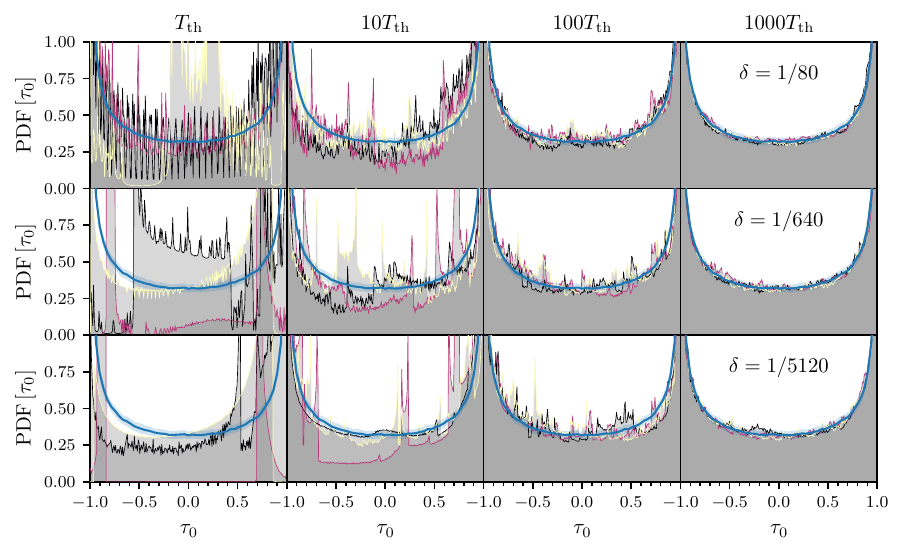}
    \caption{Ergodicity in the central spin model for $L = 46$. Each panel displays the normalized time-averaged observable $\tau_0$, computed over progressively longer time intervals (given in units of the thermalization time $\Tth$). The thermalization time $\Tth$ is determined from the spectral function, as shown in \figref{fig:centralspin_cspec}{b} of the main text. For each interval, $10^6$ equidistant measurements were performed. The blue curve corresponds to the microcanonical distribution, while the other colors represent three distinct trajectories. Columns indicate multiples of $\Tth$, while rows correspond to different values of the parameter $\delta$: $\delta = 1/80$ ($J\Tth \approx 389$), $\delta = 1/640$ ($J\Tth \approx 2954$), and $\delta = 1/5120$ ($J\Tth \approx 22780$).
    \label{fig:ergodicity}}
\end{figure*}

\paragraph*{(iii)}
The time required for a single trajectory to approximate the microcanonical ensemble significantly exceeds the thermalization time $\Tth$, as defined by the decay of the autocorrelator. While a few encounters with the chaotic manifold are sufficient to erase memory of the initial conditions, they do not guarantee adequate sampling of the full microcanonical distribution. This is demonstrated in \autoref{fig:ergodicity}, which shows the evolving distribution of $\tau_0$ for three individual trajectories (red, black, and yellow) over progressively longer evolution times (in units of $\Tth$, which depends on $\delta$) and for three distinct values of $\delta$. At each multiple of $\Tth$, the distributions exhibit qualitative similarity across the different $\delta$ values. Notably, only after $1000\Tth$ (final column) do the trajectories converge to the microcanonical distribution (blue). These results confirm our expectation that the time needed to approximate the microcanonical ensemble scales as $\Tth$ extracted from the decay of the autocorrelation, but takes significantly longer.

\section{Lower bound on the thermalization time}\label{app:lower_bound}

In this appendix, we derive a general lower bound for \(\Tth\). It follows from a simple application of the triangle inequality to bound the value of the autocorrelator given a bound on its derivative. The proof applies with minimal alteration to either classical or quantum systems. We begin with classical systems.

Consider a conserved quantity \(Q\) of an unperturbed Hamiltonian \(H_0\) (which need not be integrable), and its time evolution \(Q(t)\) under a perturbed Hamiltonian \(H = H_0 + \delta V\). We will prove that the thermalization time \(\Tth\)---the time needed for \(C_Q(t)\) to relax to \(\expval{Q}^2/\expval{Q^2}\) in a stationary state of \(H\)---is lower bounded as
\begin{equation}\label{eqn:TthBound}
    \Tth \geq \frac{1}{\vert\delta\vert}\frac{\mathrm{Var}(Q)}{\sqrt{\expval*{\poissonbracket{Q}{V}^2} \expval*{Q^2}}},
\end{equation}
where \(\mathrm{Var}(Q) = \expval{Q^2} - \expval{Q}^2\), and \(\expval{\cdots}\) is an expectation value in the stationary initial state of \(H\).

We begin by deriving a bound for the normalized autocorrelator
\begin{equation}
    C_Q(t) = \frac{\expval{Q(t)Q(0)}}{\expval{Q(0)Q(0)}},
\end{equation}
where \(Q(0) = Q\). 
We have
\begin{subequations}
\begin{align}
    \expval*{Q^2} \partial_t C_Q(t) &= \expval{\poissonbracket{Q(t)}{H}Q(0)} \\
    &= \expval{\poissonbracket{Q}{H_0 + \delta V}Q(-t)} \\
    &= \delta \expval{\poissonbracket{Q}{V}Q(-t)}. \label{eqn:dtC_expectation}
\end{align}
\end{subequations}
In the second line, we used the stationarity of the distribution to shift the time evolution to the second factor of \(Q\). In the last line, we used that \(Q\) is conserved by \(H_0\), \(\poissonbracket{Q}{H_0} = 0\).

We want to replace the expectation value in \autoref{eqn:dtC_expectation} with a simpler upper bound. The absolute value of an expectation value \(|\expval{A B}|\) can be bounded as \(\sqrt{\expval*{A^2}\expval*{B^2}}\) by the Cauchy-Schwarz inequality.

Thus, for the autocorrelator, we have
\begin{equation}\label{eqn:derivative_bound}
    |\partial_t C_Q(t)| \leq \vert\delta\vert \sqrt{\frac{\expval*{\poissonbracket{Q}{V}^2}}{\expval*{Q^2}}}=: \vert\delta\vert M,
\end{equation}
where we used that \(\expval*{Q^2} = \expval*{Q(-t)^2}\) in a stationary state. Note that the constant \(M\) does not depend on the system size when \(V\)  and \(Q\) are either local operators or extensive sums of local operators.

The bound on the derivative \autoref{eqn:derivative_bound} immediately translates to a bound on the decay of the autocorrelator,
\begin{subequations}
\begin{align}
    |C_Q(t) - 1| &= \left| \int_0^t \partial_t C_Q(t') \mathrm{d}t' \right| \\
    &\leq \int_0^t |\partial_t C_Q(t')| \mathrm{d}t' \\
    &\leq \vert\delta\vert M t.
\end{align}
\end{subequations}
We take \(t \geq  0\).

This bound on the autocorrelator can be reinterpreted as a bound on the thermalization time \(\Tth\). Say we define \(\Tth\) as the smallest time such that \(C_Q(t)\) remains within some small window \(\epsilon>0\) of its equilibrium value \(C_Q(\infty) = \expval{Q}^2 / \expval{Q^2} \leq 1\) for all \(t \geq \Tth\),
\begin{equation}
    \left|C_Q(t) - C_Q(\infty)\right| \leq \epsilon 
    \quad\text{for all }t \geq \Tth \geq 0.
\end{equation}
The reverse triangle inequality allows us to relate this condition to our bound on \(C_Q(t)\),
\begin{equation}
    \left||1-C_Q(\infty)| - |C_Q(t) - C_Q(\infty)|\right| \leq |C_Q(t) - 1| \leq \vert\delta\vert M t.
\end{equation}
Setting \(t = \Tth\) and \(|C_Q(t) - C_Q(\infty)| = \epsilon\), we have
\begin{equation}
    |1- C_Q(\infty) - \epsilon| \leq \vert\delta\vert M \Tth.
\end{equation}
Reinstating the explicit expressions in terms of \(Q\) and \(V\), we arrive at a bound on \(\Tth\),
\begin{equation}
    \Tth \geq \frac{1}{\vert\delta\vert}\frac{|(1-\epsilon)\expval{Q^2} - \expval{Q}^2|}{\sqrt{\expval*{\poissonbracket{Q}{V}^2} \expval*{Q^2}}}.
\end{equation}
The case of interest is \(\epsilon \ll 1\). Our bound is finite as \(\epsilon \to 0\), so for small enough \(\epsilon\) we must have
\begin{equation}
    \Tth \geq \frac{1}{\vert\delta\vert}\frac{\mathrm{Var}(Q)}{\sqrt{\expval*{\poissonbracket{Q}{V}^2} \expval*{Q^2}}},
\end{equation}
which is \autoref{eqn:TthBound}.

We note that our bound is properly a bound on the time required for \(C_Q(t)\) to first cross its equilibrium value. There may be a tighter bound which properly incorporates the condition that \(C_Q(t)\) must remain close to its equilibrium value for all later times. However, we believe the scaling with \(\delta\) in this bound is already optimal.

A subtlety in the application of this bound to the central spin model is that, because only the \(E=0\) manifold in \(H_0\) is integrable, for most of the conserved quantities we do not have \(\expval{\poissonbracket{Q}{H_0}} = 0\) in stationary states of \(H = H_0 + \delta V\). Indeed, for the observable \(\tau_0\) studied in the main text, we have \(\poissonbracket{\tau_0}{V} = 0\), as these operators have disjoint support. The non-conservation of \(\tau_0\) relies on the microcanonical shell deforming to include parts of phase space where \(\expval*{\poissonbracket{\tau_0}{H_0}} \neq 0\). In this case, it is appropriate to skip the step in our derivation where we set \(\expval{\poissonbracket{Q}{H_0}} = 0\), and use a more general bound
\begin{equation}
    \Tth \geq \frac{\mathrm{Var}(Q)}{\sqrt{\expval*{\poissonbracket{Q}{H}^2} \expval*{Q^2}}}.
\end{equation}
For the central spin model, we have that \(|\poissonbracket{\tau_0}{H}| = |\delta| W\), where \(W\) is a singular function on the \(E=0\) microcanonical shell. Numerically, this singular function appears to have a finite integral as \(\delta \to 0\), but we cannot rigorously confirm that the resulting bound on \(\Tth\) is \(\order{\vert\delta\vert^{-1}}\) in this case. The bound will also typically not be \(\order{1}\) with system size, because the rate of evolution of an observable supported on the central site is generically \(\order{L}\) with our choice of normalization of \(H\).

Another straightforward generalization of this bound is to use the more general H\"older inequality rather than the Cauchy-Schwarz inequality. We have \(|\expval{AB}| \leq \expval*{|A|^p}^{1/p} \expval*{|B|^q}^{1/q}\) for any \(p,q\geq 1\) such that \(1/p + 1/q = 1\). Thus, we have
\begin{equation}\label{eqn:TthBound_pq}
    \Tth \geq \frac{\mathrm{Var}(Q)}{\expval*{|\poissonbracket{Q}{H}|^p}^{1/p} \expval*{|Q|^q}^{1/q}},
\end{equation}
for any such \(p\) and \(q\). The scaling of \(\Tth\) with \(\delta\) is the same in this version of the bound, but the quantitative bound may be improved by taking \(p \neq 2\) (which reproduces the Cauchy-Schwarz inequality) for certain instances of \(Q\) and \(H\). For instance, taking \(p=1\), \(q=\infty\), we have
\begin{equation}
    \Tth \geq \frac{\mathrm{Var}(Q)}{\expval*{|\poissonbracket{Q}{H}|} \|Q\|_\infty},
\end{equation}
where \(\|Q\|_\infty\) is the essential supremum of \(|Q|\) in the steady state defining the autocorrelator.

Finally, we provide the quantum generalization of these bounds. All steps of the proof follow through identically upon replacing Poisson brackets with commutators (we set \(\hbar=1\)), except for the H\"older inequality. This step becomes more  complicated due to non-commutativity. For general complex matrices \(A\) and \(B\) and a quantum state \(\rho\) we have
\begin{equation}
    \expval*{A^\dagger B} = \Tr[A^\dagger B \rho] = \Tr[\rho^r A^\dagger B \rho^{1-r}],
\end{equation}
where \(r \in [0,1]\). The H\"older inequality for the trace gives
\begin{multline}
    |\expval*{A^\dagger B}| \leq \left|\Tr[(\rho^r A^\dagger A \rho^{r})^{p/2}]\right|^{1/p} \\
    \times\left|\Tr[(\rho^{1-r} B^\dagger B \rho^{1-r})^{q/2}]\right|^{1/q},
\end{multline}
for any \(p,q \geq 1\) such that \(1/p + 1/q = 1\). This factor replaces \(\expval*{|A|^p}^{1/p} \expval*{|B|^q}^{1/q}\) in \autoref{eqn:TthBound_pq}. Particular cases of interest are \(r=p=1\), \(q=\infty\),
\begin{equation}
    \Tth \geq \frac{\mathrm{Var}(Q)}{\Tr[\sqrt{\rho |\comm{Q}{H}|^2 \rho}] \|Q\|_\infty},
\end{equation}
(where now \(\|Q\|_\infty\) is the maximum singular value of \(Q\)), and the Cauchy-Schwarz case of \(r=1/2\) and \(p=q=2\),
\begin{equation}
    \Tth \geq \frac{\mathrm{Var}(Q)}{\sqrt{\expval*{|\comm{Q}{H}|^2} \expval*{Q^2}}}.
\end{equation}

\section{Correlators and spectral functions}\label{app:corr_spec}

\begin{figure*}[!htbp]
    \centering
    \includegraphics[width=\textwidth]{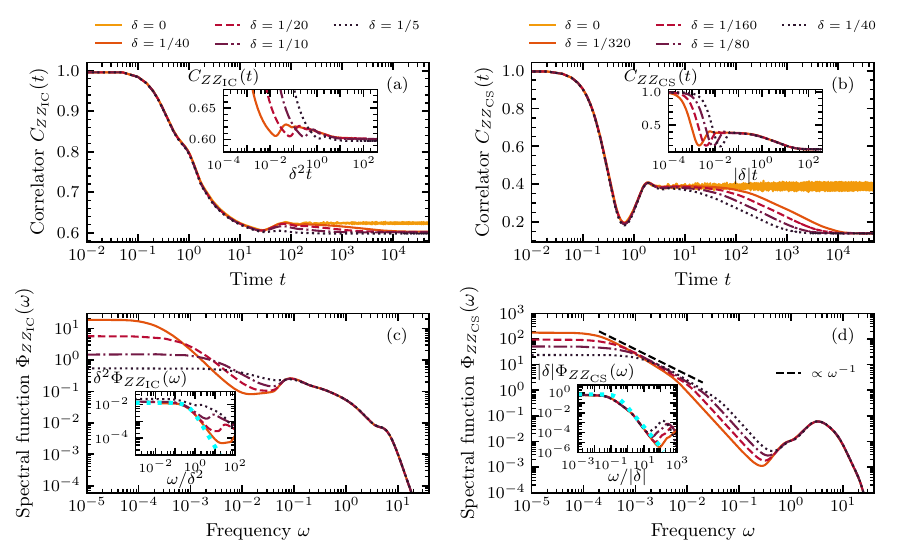}%
    \caption[]{Autocorrelator and spectral function for varying perturbation strength $\delta$ of non-conserved observables $ZZ_\mathrm{IC}$ and $ZZ_\mathrm{CS}$ defined in \autoref{eq:non_conserved_observable} for the Ishimori spin chain and central spin model, respectively. Panels (a) and (b) show the correlators $C_{ZZ_\mathrm{IC}}(t)$ and $C_{ZZ_\mathrm{CS}}(t)$, while their insets depict the scaling collapses upon rescaling time. Note that the $y$-axis is reduced compared to the plots in the main text, as the autocorrelator does not decay to zero. Panels (c) and (d) show the spectral functions $\Phi_{ZZ_\mathrm{IC}}(\omega)$ and $\Phi_{ZZ_\mathrm{CS}}(\omega)$, respectively. In panel (d), the dashed line indicates the $\omega^{-1}$ scaling. We observe scaling collapse in the insets by rescaling. The blue dotted lines indicate approximate Lorentzian fits (up to some scalar constant). These results for the non-conserved observables align with our main results that $\Tmelt, \Tth \propto \delta^{-2}$ for the Ishimori spin chain and $\Tmelt, \Tth \propto \vert\delta\vert^{-1}$ for the central spin model. Parameters used: $L = 45$ for the Ishimori chain and $L = 46$ for the central spin model, $E = 0$, and $\Sztot = 0$ for both.
    \label{fig:corr_spec_non_conserved}}
\end{figure*}

In this section, we explain how the autocorrelators and their spectral functions are numerically computed. We also detail the procedure used to suppress statistical noise in these quantities while maintaining computational efficiency.

A naive computation of the correlator using \autoref{eq:correlation} results in significant errors due to finite sampling of the microcanonical shell. While these errors can be suppressed by sampling more trajectories, there exists a simpler alternative. As done in Ref.~\cite{lim2024defining}, we consider the time-averaged quantity for the correlator:
\begin{equation}
    \overline{\expval{O(t)O(0)}} =\lim\limits_{T \to \infty} \frac{1}{2T} \int^T_{-T} \dd{\tau} \expval{O(\tau+t)O(\tau)}, \label{eq:corr_time_averaged}
\end{equation}
where $T$ is the evolution time.
This quantity $\overline{\expval{O(t)O(0)}}$ time-averages the correlators with fixed time difference $t$ in a window of width $2T$ about $\tau = 0$. It is also simple and efficient to compute. To show this, consider the Fourier transform of this quantity---the spectral function. Define a window function $w_T(t)$, which is unity for $t \in [-T,T]$ and zero otherwise, and the windowed observable $O_T(t) = w_T(t) O(t)$. Then, the spectral function is given by the following:
\begin{align}
    \Phi_O(\omega) &= \lim\limits_{T \to \infty} \frac{1}{4\pi T} \int^\infty_{-\infty} \dd{t} e^{i\omega t} \int^T_{-T} \dd{\tau} \expval{O(\tau+t)O(\tau)} \nonumber \\
    &\approx \frac{1}{4\pi T} \int^\infty_{-\infty} \dd{t} e^{i\omega t} \int^\infty_{-\infty} \dd{\tau} \expval{O_T(\tau+t)O_T(\tau)} \nonumber \\\
    &= \frac{\pi}{T} \expval{\hat{O}_T(\omega) \hat{O}_T(-\omega)} = \frac{\pi}{T} \expval{\abs*{\hat{O}_T(\omega)}^2}, \label{eq:spec_time_averaged}
\end{align}
where we approximate the $T \to \infty$ limit with a sufficiently large $T$. The first equality on the last line is due to the convolution theorem while the the second equality is a consequence of $O(t)$ being a real-valued function. Here, $\hat{O}_T(\omega)$ refers to the Fourier transform of the windowed observable. 

From \autoref{eq:spec_time_averaged}, it is evident how this method has better convergence properties: the contribution from each trajectory to the spectral function from $\abs*{\hat{O}_T(\omega)}^2$ is manifestly positive while that from $\expval{O(t)O(0)}$ in \autoref{eq:spectral_function} in the main text is not necessarily positive.  Then, the time-averaged correlator $\overline{\expval{O(t)O(0)}}$ can be computed by taking the inverse Fourier transform of \autoref{eq:spec_time_averaged}. To compute \autoref{eq:spec_time_averaged}, we first compute $O(t)$ for $-T < t < T$ by evolving forward and backward in time using two copies of the initial trajectory, and then we compute $\hat{O}_T(\omega)$ by performing a fast Fourier transform (FFT) operation on $O(t)$. Finally, we simply compute the phase-space average to obtain $\expval*{\abs*{\hat{O}_T(\omega)}^2}$. Indeed this method of finding $\overline{\expval{O(t)O(0)}}$ is quite efficient because performing FFT takes $\mathcal{O}(N \log{N})$ time, where $N$ is the number of time steps in the evolution, while a naive computation of $\overline{\expval{O(t)O(0)}}$ using \autoref{eq:corr_time_averaged} is an $\mathcal{O}(N^2)$ operation.

\section{Dynamics of other observables} \label{app:non_conserved_observables}

In this appendix, we consider non-conserved observables to show that the scalings of our relevant timescales ($\Tphi$, $\Tlya$, $\Tmelt$, and $\Tth$) are identical even for non-conserved quantities. Namely:
\begin{equation}
    ZZ_\mathrm{IC} = \sum^{L-1}_{j=0} S_j^z S_{j+1}^z \quad \text{and} \quad ZZ_\mathrm{CS} = \sum^{L-1}_{j=1} g_j S_0^z S_j^z,
     \label{eq:non_conserved_observable}
\end{equation}
for the Ishimori spin chain and central spin model, respectively.

In \figref{fig:corr_spec_non_conserved}{a} and \figref{fig:corr_spec_non_conserved}{b} (\figref{fig:corr_spec_non_conserved}{c} and \figref{fig:corr_spec_non_conserved}{d}), we plot the correlators (spectral functions) of $ZZ_\mathrm{IC}$ and $ZZ_\mathrm{CS}$, respectively. First, we numerically confirm that these observables are not conserved in the integrable limit ($\delta = 0$) by observing that the correlator $C_{ZZ_\mathrm{IC}}(t)$ ($C_{ZZ_\mathrm{CS}}(t)$) decays from the initial value as shown by the orange line in \figref{fig:corr_spec_non_conserved}{a} (\figref{fig:corr_spec_non_conserved}{b}). For $\abs{\delta}>0$, the correlators thus decay to the plateau reached by the orange line on the scale $\Tphi$. We find that $\Tphi \propto \delta^0$. Notably, both $C_{ZZ_\mathrm{IC}}(t)$ and $C_{ZZ_\mathrm{CS}}(t)$ decay to nonzero values, consistent with the presence of large Drude weights in the corresponding spectral functions. We remark that $\Tphi$ occurs at a much later time than for the observables studied in the main text (for the same value of $\delta$).

\begin{figure}[!t]
    \centering
    \includegraphics[width=\columnwidth]{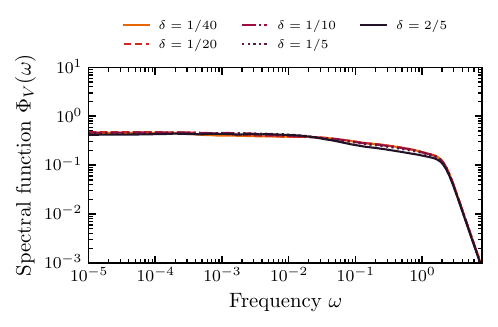}%
    \caption[]{Spectral function of $V$ with varying perturbation strengths for the Ishimori chain (with $L = 45$). The data shows saturation of $\Phi_{V}(\omega)$ for all values of $\delta$ as the perturbation $V$ has zero Drude weight.
    \label{fig:ishimori_spec_V}}
\end{figure}

Nevertheless, we observe the same scaling collapses of the correlators (at late times) and their respective spectral functions (at low frequencies): $\Tmelt, \Tth \propto \delta^{-2}$ for the Ishimori spin chain and $\Tmelt, \Tth \propto \abs{\delta}^{-1}$ for the central spin model. We also find Lorentzian shapes of the spectral functions in the Ishimori chain. 

However, in the central spin model, the spectral function is now slightly modified. Instead of the expected $\omega^{-2}$ scaling throughout the whole region of $\omega \gtrsim \Gamma$, we observe an intermediate region that scales as $\omega^{-1}$ before saturating to a constant at a lower frequency (see Lorentzian fit shown as blue dotted line in the inset of  \figref{fig:corr_spec_non_conserved}{d}). Note that we also see this for $\Phi_V(\omega)$ in \autoref{fig:central_spin_spec_V} in the main text. The inset of \figref{fig:corr_spec_non_conserved}{d} shows that this $\omega^{-1}$ region has the same, fixed frequency range across different $\delta$'s.

Finally, we conclude this section by computing the spectral function $\Phi_V(\omega)$ for the Ishimori spin chain in \autoref{fig:ishimori_spec_V}, complementing \autoref{fig:central_spin_spec_V} in the main text. In contrast to $\Phi_V(\omega)$ in the central spin model, here the Drude weight is zero as we observe no broadening of the spectral function as $\delta$ increases. This aligns with our main results that the timescales in the Ishimori spin chain follow standard perturbation theory since $\Phi_V(0) < \infty$. 

\begin{figure}[!htbp]
    \centering
    \includegraphics[width=\columnwidth]{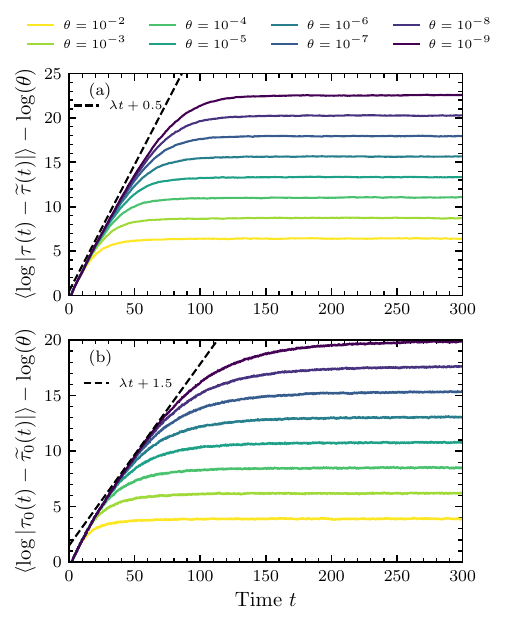}%
    \caption[]{Lyapunov exponent measured by an out-of-time ordered correlator (OTOC). Given an initial spin trajectory $\vb{S}$, we find its perturbed counterpart $\widetilde{\vb{S}}$ by rotating each spin of $\vb{S}$ along the $x$-axis by angle $\theta$. Using $\vb{S}(t)$ and $\widetilde{\vb{S}}(t)$, we measure $O(t)$ and $\widetilde{O}(t)$ and compute the difference $\abs*{O(t) - \widetilde{O}(t)}$ for different values of $\theta$. Panel (a) refers to the Ishimori chain with $L = 45$ and $\delta = 1/5$ ($O = \tau$ from \autoref{eq:torsion} in the main text) while panel (b) refers to the central spin model with $L = 46$ and $\delta = 1/5$ ($O = \tau_0$ from \autoref{eq:centralspin_observable} in the main text). The exponential growth rate of $\abs*{O(t) - \widetilde{O}(t)}$ agrees with $\lambda t$ as shown by the dashed line in each plot, where $\lambda$ is numerically computed using ChaosTools.jl~\cite{ChaosTools}, which is based on Ref.~\cite{benettin1976kolmogorov}. 
    \label{fig:otoc}}
\end{figure}

\section{Lyapunov exponents and out-of-time-correlators} \label{app:lyapunov_otocs}
Here, we provide provide an alternative approach for extracting Lyapunov exponents using the out-of-time-correlator (OTOC), which agrees with the Lyapunov exponents computed using the divergence of trajectories.

The Lyapunov exponent can also be measured via the out-of-time ordered correlator~\cite{rozenbaum_lyapunov_2017} in both classical and quantum systems (in the semi-classical regime). The OTOC can be defined for observables $A$ and $B$ and is measured as $\mathrm{OTOC} = \expval*{\poissonbracket{A(t)}{B}^2}$. Prior work has shown that the growth rate of the OTOC is set by twice the Lyapunov exponent~\cite{rozenbaum_lyapunov_2017}. Equivalently, we can also compute $\abs*{O(t) - \widetilde{O}(t)}$, where we use two nearby initial spin states $\vb{S}$ and $\widetilde{\vb{S}}$ to measure $O$ and $\widetilde{O}$, respectively. The state $\widetilde{\vb{S}}$ is initialized by rotating each spin of $\vb{S}$ about the $x$-axis by some angle $\theta$. This quantity---the difference between the observable and its perturbed counterpart---is equivalent to the OTOC up to leading order in $\theta$. To observe this, let's denote the generator of our rotation as $R = \sum_i S_i^x$ such that $\widetilde{O}(0) \approx O(0) + \theta \poissonbracket{O(0)}{R}$. Since $H$ is the generator for time evolution, where
\begin{equation}
    \widetilde{O}(t) = \widetilde{O}(0) + t \pob{\widetilde{O}(0)}{H} + \frac{t^2}{2!} \pob{\pob{\widetilde{O}(0)}{H}}{H} + \ldots,
\end{equation}
we can show that (up to first order in $\theta$), 
\begin{align}
\begin{split}
    \abs*{O(t)-\widetilde{O}(t)} \approx \abs{\theta} &\big\vert\pob{O(0)}{R} + t \pob{\pob{O(0)}{R}}{H} \\
    & \quad + \frac{t^2}{2!} \pob{\pob{\pob{O(0)}{R}}{H}}{H} + \ldots \big\vert.
\end{split}
\end{align}
In the case that $\pob{R}{H} = 0$, it is straightforward to show that $\abs*{O(t) - \widetilde{O}(t)} \approx \abs{\theta}\abs{\pob{O(t)}{R}} \propto \abs{\theta} e^{\lambda t}$. Thus, taking the square before phase-space averaging yields $\mathrm{OTOC} = \abs{\theta}^2 \expval*{\abs{\pob{O(t)}{R}}^2} \propto \abs{\theta}^2 e^{2 \lambda t}$. On the other hand, if $\pob{R}{H} \neq 0$ (which is true for both our models), then it can be shown that (up to $\mathcal{O}(t^2)$ corrections)
\begin{equation}
    \abs*{O(t) - \widetilde{O}(t)} \approx \abs{\theta}\abs{\pob{O(t)}{R} + \pob{R(t)}{O(0)}}, 
\end{equation}
which still scales as $\abs{\theta} e^{\lambda t}$ if we assume that each term of the summation scales as $e^{\lambda t}$.

We measure this quantity $\abs*{O(t) - \widetilde{O}(t)}$ for different values of $\theta$ and find the scaling relationship $\abs*{O(t) - \widetilde{O}(t)} \propto \abs{\theta} e^{\lambda t}$ for sufficiently small values of $\theta$ (measured in radians). In \autoref{fig:otoc}, we measure $\tau$ (for the Ishimori chain) and $\tau_0$ (central spin model), which are defined in the main text, and compute $\expval*{\log{\abs*{O(t) - \widetilde{O}(t)}}}$. Note that in \autoref{fig:otoc} we take the phase space average after applying the logarithm. This ensures that we are comparing against $\expval{\lambda}t$. As expected, $\abs*{O(t) - \widetilde{O}(t)}$ grows exponentially with its rate set by the Lyapunov exponent $\lambda$. 

\section{System-size dependencies of timescales} \label{app:finite_size}

In this section, we consider how the timescales ($\Tlya$, $\Tmelt$, and $\Tth$) vary with system sizes at a fixed perturbation strength. This complements our main results that mostly dealt with scaling relationships with varying perturbation strengths at fixed system sizes. In summary, we find that $\Tlya, \Tmelt \propto 1/\log(L)$ while $\Tth \propto L^0$ in both of our models using the conserved observable $\tau$ and $\tau_0$.

\begin{figure}[!b]
    \centering
    \includegraphics[width=\columnwidth]{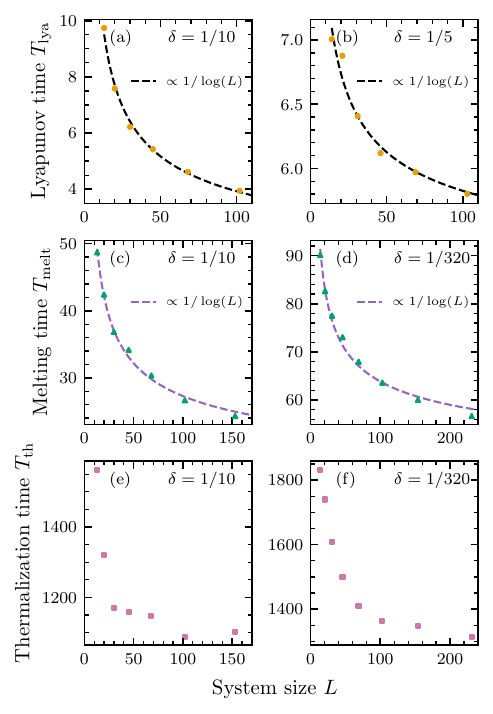}
    \caption[]{System-size dependencies of timescales. Panels (a,c,e) refer to the Ishimori spin chain, while (b,d,f) refer to the central spin model. (a) and (b) present the finite-size effects of the Lyapunov time $\Tlya$. The dashed black lines indicate the $1/\log(L)$ fits. (c) and (d) show the melting time $\Tmelt$ while (e) and (f) show the thermalization time $\Tth$; these were obtained using correlators and spectral functions of the same observables used in the main text (torsion $\tau$ for Ishimori spin chain and $\tau_0$ for central spin model). In (c) and (d), the dashed blue lines show the $1/\log(L)$ fits. The $\delta$ used for each plot is shown in its respective legend.} \label{fig:system_size_dependence}
\end{figure} 

\begin{figure}[!htbp]
    \centering
    \includegraphics[width=\columnwidth]{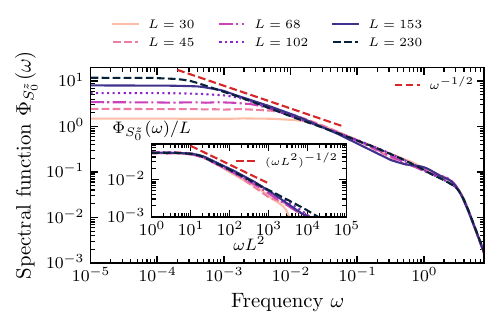}
    \caption[]{Evidence for diffusion of $S_0^z$ in the Ishimori spin chain. We compute the spectral function of the local observable $S_0^z$ in the Ishimori spin chain for different system sizes at a fixed $\delta = 1$. As $L$ grows, we observe a growing diffusive regime with ${\omega}^{-1/2}$ scaling, as indicated by the red dashed line. The inset illustrates the scaling collapse after rescaling $\Phi_{S_0^z}(\omega)/L$ against $\omega L^2$. Note that we choose $\delta=1$ to access the diffusive regime for the system sizes shown here.} \label{fig:diffusion}
\end{figure}

In (a) and (b) of \autoref{fig:system_size_dependence}, we compute $\lambda$ (and thereby Lyapunov time $\Tlya = 1/\lambda$) for various system sizes in both models and find that $\lambda \propto \log(L)$, as indicated by the dashed black lines. This logarithmic divergence is likely due to the order of limits taken. If we take $L \to \infty$ first and then $T \to \infty$, where $T$ is the simulation time, then we expect to observe no logarithmic divergence of $\lambda$. In our work, we take $T \to \infty$ first and then $L \to \infty$.

Now, we focus on the finite-size effects of melting time $\Tmelt$ and thermalization time $\Tth$, shown in (c) and (e) of \autoref{fig:system_size_dependence} for the Ishimori spin chain and (d) and (f) of \autoref{fig:system_size_dependence} for central spin model, respectively. Our data suggests that $\Tth$ approximately saturates to a constant value at sufficiently large $L$ while $\Tmelt$ keeps slowly decreasing with $L$ and in the range of system sizes we analyzed can be approximated by $\Tmelt \propto 1/\log(L)$ as shown by the dashed blue lines in (c) and (d) of \autoref{fig:system_size_dependence}. We, however, do not know if this trend continues for larger $L$ or if $\Tmelt$ also saturates. Interestingly, despite the fact that one model (Ishimori) is local and the other (central spin) is infinite range, they both have very similar weak scaling of relevant timescales with the system size. For the central spin model, independence of the thermalization time on $L$ is expected due to the lack of spatial locality. For the Ishimori chain, as explained in \autoref{ssec:ishimori_systemsize} in the main text, thermalization time is also expected to be $L$-independent because the torsion $\tau$ does not overlap with either conserved quantities $\HIC$ or $\Sztot$. The situation changes for the Ishimori chain if we consider a local observable that has non-vanishing overlap with a conserved quantity in the perturbed model, such as $S_0^z$ at strong integrability breaking. Then, as noted in \autoref{ssec:ishimori_systemsize}, we observe system size dependence of thermalization time $\Tth \propto L^2$ consistent with diffusion. This is shown in \autoref{fig:diffusion}.

\end{document}